\documentclass[aps,prd,amsmath
,eqsecnum 
,preprintnumbers
,nofootinbib 
 ,showpacs 
 ,showkeys 
,twocolumn 
]{revtex4}
\usepackage[dvips]{graphicx}
\usepackage{amsmath}
\usepackage{amsfonts}
\usepackage{amssymb}
\usepackage{longtable} 
\usepackage{color}

\allowdisplaybreaks[1] 

\newcommand{\df}{\ {\overset {\rm def} =}\ }
\newcommand{\dr}[2]{\frac {{\rm d} {#1}} {{\rm d} {#2}}}

\newcommand{\dril}[2]{{{\rm d} {#1}} / {{\rm d} {#2}}}

\newcommand{\llim}[1] {\ {\underset {#1} {\longrightarrow}}\ }

\begin{document}

\title{Properties of blueshifted light rays in quasispherical Szekeres
metrics}

\author{Andrzej Krasi\'nski}
\affiliation{N. Copernicus Astronomical Centre, Polish Academy of Sciences, \\
Bartycka 18, 00 716 Warszawa, Poland} \email{akr@camk.edu.pl}

\date {}

\begin{abstract}
This paper is a follow-up on two previous ones, in which properties of
blueshifted rays were investigated in Lema\^{\i}tre -- Tolman (L--T) and
quasispherical Szekeres (QSS) spacetimes. In the present paper, an axially
symmetric QSS deformation is superposed on such an L--T background that was
proved, in the first paper, to mimic several properties of gamma-ray bursts. The
present model makes $z$ closer to $-1$ than in the background L--T spacetime,
and, as implied by the second paper, strong blueshifts exist in it only along
two opposite directions. The QSS region is matched into a Friedmann background.
The Big Bang (BB) function $t_B(r)$, which is constant in the Friedmann region,
has a gate-shaped hump in the QSS region. Since a QSS island generates stronger
blueshifts than an L--T island, the BB hump can be made lower -- then it is
further removed from the observer and implies a smaller observed angular radius
of the source. Consequently, more sources can be fitted into the sky -- all
these facts are confirmed by numerical computations. Null geodesics reaching
present observers from different directions relative to the BB hump are
numerically calculated. Patterns of redshift across the image of the source and
along the rays are displayed.
\end{abstract}

\maketitle

\section{Motivation and background}\label{intro}

\setcounter{equation}{0}

In Lema\^{\i}tre \cite{Lema1933} -- Tolman \cite{Tolm1934} and Szekeres
\cite{Szek1975,Szek1975b} spacetimes, some of the light rays emitted at the Big
Bang (BB) reach all observers with infinite {\em blue}shift ($1 + z \df
\nu_e/\nu_o = 0$, where $\nu_e$ and $\nu_o$ are frequencies of the emitted and
observed radiation, respectively). This is in contrast to Robertson -- Walker
spacetimes, where all light from the BB is observed with $z = \infty$
\cite{Elli1971,PlKr2006}. The quantity $z$, traditionally called {\em red}shift,
being negative (and then called blueshift) means that the frequency observed is
higher than the frequency at the emission point, and $z \to -1$ implies $\nu_o
\to \infty$. The existence of blueshifts in L--T models was predicted by
Szekeres in 1980 \cite{Szek1980}, in a casual remark without proof, and then
confirmed by Hellaby and Lake in 1984 \cite{HeLa1984} by explicit calculation.

Two conditions are necessary for infinite blueshift:

(1) The BB time at the emission point of the ray must have nonzero spatial
derivative in comoving-synchronous coordinates (the BB is ``nonsimultaneous'').

(2) The ray is emitted at the BB in a radial direction.

\noindent Condition (2) was derived in Ref. \cite{HeLa1984}, but seems to have
been overlooked by all later authors until Ref. \cite{Kras2016a}, even though it
follows quite simply from the geodesic equations. The two conditions together
seem to be also sufficient, but a general proof of their sufficiency still does
not exist; it is only implied by the full list of separate cases \cite{HeLa1984}
and hinted at by numerical calculations \cite{Kras2016a,Kras2014d}.

The Szekeres spacetimes \cite{Szek1975,Szek1975b}, in general, have no symmetry,
thus no radial directions. In view of condition (2) it was not clear whether any
rays with infinite blueshift exist in them. This question was addressed in Ref.
\cite{Kras2016b}. It was shown that in an axially symmetric quasispherical
Szekeres (QSS) spacetime, $z = -1$ can possibly happen on axial rays; i.e.,
those that intersect every space of constant time on the symmetry axis. It was
then confirmed by a numerical calculation in an exemplary QSS model that $1 + z
< 10^{-5}$ along axial rays emitted from the BB. It was also shown, by a blind
numerical search, that rays with $1 + z < 0.07$, and with similar spatial
profiles of $z$ along neighbouring rays, exist in an exemplary fully
nonsymmetric QSS model.

Since the L--T and Szekeres models have been proven to successfully describe
several observed features of our Universe \cite{BoCK2011,SuGa2015}, and they
predict a possible existence of blueshifts, one must thoroughly test the
implications of blueshifts in order to either find a place for them among the
observed phenomena, or conclude that the BB in the real Universe must have been
simultaneous. With this motivation, it was shown in Ref. \cite{Kras2016a} that
an L--T region with a gate-shaped ``hump'' on the BB profile matched into a
Friedmann background can mimic some observed properties of gamma-ray bursts
(GRBs), such as the frequency range ($0.24 \times 10^{19}$ to $1.25 \times
10^{23}$ Hz), the existence of afterglows and the large distances to the
sources. Placing several different L--T regions in the same Friedmann background
would then account for the large number of possible sources. However, the model
of Ref. \cite{Kras2016a} was unsuccessful on two accounts:

(1) The gamma-ray flashes and the afterglows lasted for too long. The model
contains a parameter that should allow for controlling the durations, but
insufficient numerical accuracy did not permit actual use of it.

(2) The radiation was emitted isotropically instead of being collimated into
narrow beams, as the observed GRBs are supposed to be \cite{Perlwww}.

Also, the model of Ref. \cite{Kras2016a} left some problems open. The main one
was: how small could the humps on the BB profile be made while still generating
the right range of frequencies of the observed radiation.\footnote{It is easy to
obtain small $1 + z$ with a high hump on the BB, but then the radiation source
is close to the observer and has a large angular diameter in the sky. With a
lower hump the diameter gets smaller, but $1 + z$ gets larger. Keeping both the
diameter and $1 + z$ sufficiently small is the main difficulty.}

Ref. \cite{Kras2016b} was the first step in improving the model of Ref.
\cite{Kras2016a}. It showed by examples that strongly blueshifted rays in QSS
spacetimes exist only along two opposite directions. That paper also proved that
in a QSS model the minimum $1 + z$ is smaller than in an L--T model that has the
same BB profile.

The present paper builds upon this last observation. The model considered here
is a QSS deformation superposed on the L--T region of Ref. \cite{Kras2016a}.
Since the QSS deformation results in a smaller $1 + z$ at the observer, the
minimum value of $1 + z$ found in Ref. \cite{Kras2016a} can be achieved with a
lower BB hump. This implies a greater distance between the source of radiation
and the observer, and a smaller angular diameter of the source seen in the sky.
The progress achieved with respect to Ref. \cite{Kras2016a} is rather moderate,
but this cannot be the ultimate limit of improvement: the class of BB profiles
used here was found by trial and error (see Sec. \ref{impro}), and it is
impossible that the optimal shape could be hit upon in this way.

The L--T and Szekeres metrics are solutions of the Einstein equations with a
dust source, so they cannot apply to the real Universe at such early times when
pressure cannot be neglected. It is assumed that they may apply onward from the
end of the last-scattering (LS) epoch. The mean mass density at LS, denoted
$\rho_{\rm LS}$, in the now-standard $\Lambda$CDM model is known
\cite{Kras2016a}, see Sec. \ref{backgrmodel}. For every past-directed null
geodesic in a QSS (or L--T) region, the mass density at the running point is
numerically calculated. When this density becomes equal to $\rho_{\rm LS}$, the
integration is stopped. Thus, $1 + z$ between LS and the present time is bounded
from below, $z_{\rm LS} \geq z_{\rm min} > -1$. The computational problem is to
arrange the BB profile so that it makes $z_{\rm LS}$ sufficiently near to $-1$
($1 + z_{\rm LS} < 1.689 \times 10^{-5}$ \cite{Kras2016a}), but does not lead to
perturbations of the CMB radiation larger than observations allow. Among other
things, this implies that the model must be capable of making the angular
diameter of the radiation sources smaller than the observed diameter of the GRBs
(currently\footnote{\label{Linda}Private communication in 2015 from Linda
Sparke, then at NASA. The $1^{\circ}$ is the current resolution of the detectors
rather than the true diameter.} $\approx 1^{\circ}$, see Sec. \ref{inthesky}).

In Secs. \ref{QSSS} and \ref{QSShere}, the subfamily of QSS models employed here
is presented. It is an axially symmetric QSS region matched into a Friedmann
background with curvature index $k = -0.4$. In Sec. \ref{backgrmodel} the
parameters of the background model are specified. They are different from those
of the $\Lambda$CDM model \cite{Plan2014,Plan2014b} -- it was convenient to keep
them the same as in the earlier papers by this author
\cite{Kras2016a,Kras2014d}. In Sec. \ref{symmetric}, the equations of null
geodesics in the QSS region are presented. In Sec. \ref{axreds}, basic
properties of redshift are described, and the conditions for $z = -1$ in an
axially symmetric QSS model are spelled out. In Sec. \ref{ERS}, the equation of
the Extremum Redshift Surface (ERS) is derived,\footnote{Sections \ref{QSSS},
\ref{backgrmodel}, \ref{symmetric} and \ref{ERS} are partly copied from Ref.
\cite{Kras2016b}.} on which $z$ has maxima or minima along axial rays. In Sec.
\ref{corrM2}, the numerical parameters of the model used here are adapted to the
GRBs of lowest frequency. In Sec. \ref{noax}, exemplary nonaxial plane rays
reaching the present observers are numerically determined. The observers are
placed in three directions with respect to the QSS region: (I) -- in
prolongation of the dipole minimum, (II) -- in prolongation of the dipole
maximum, and (III) -- in prolongation of the dipole equator of the boundary of
the QSS region. For each observer, the redshift profiles {\em across} the image
of the radiation source are presented in tables. In Sec. \ref{noaxreds},
redshift profiles {\em along} the nonaxial rays reaching Observer I are
displayed to show that analogues of the ERS exist also along nonaxial
directions. In Sec. \ref{inthesky} it is estimated that $\approx 11,000$
radiation sources of Sec. \ref{corrM2} could be fitted into the celestial
sphere. The necessary and possible improvements of the model are discussed in
Sec. \ref{impro}. Section \ref{sumup} contains the summary and conclusions.

The present paper is a study in the geometry of the QSS spacetimes and in
properties of their blueshifted rays. Also, it introduces methods that can be
used in further refinements of the model. The observed parameters of the GRBs
were used as a beacon pointing the way, but the configuration derived here needs
further improvements before it can be considered a model of a GRB source; see
Sec. \ref{inthesky}.

Most results of numerical calculations are quoted up to 17 decimal digits. Such
precision is needed to capture time intervals of $\approx 10$ min at the
observer, which is $\approx 2 \times 10^{-16}$ in the units used here, see Sec.
\ref{QSShere}. (The 10 min is a representative time during which GRBs are
visible to the detectors \cite{Kras2016a}.)

\section{QSS spacetimes}\label{QSSS}

\setcounter{equation}{0}

The metric of the QSS spacetimes is \cite{Szek1975,Szek1975b,PlKr2006,Hell1996}
\begin{equation}\label{2.1}
{\rm d} s^2 = {\rm d} t^2 - \frac {\left(\Phi,_r - \Phi {\cal E},_r/{\cal
E}\right)^2} {1 + 2 E(r)} {\rm d} r^2 - \left(\frac {\Phi} {\cal E}\right)^2
\left({\rm d} x^2 + {\rm d} y^2\right),\ \ \ \ \
\end{equation}
\begin{equation}\label{2.2}
{\cal E} \df \frac S 2 \left[\left(\frac {x - P} S\right)^2 + \left(\frac {y -
Q} S\right)^2 + 1\right],
\end{equation}
$P(r)$, $Q(r)$, $S(r)$ and $E(r)$ being arbitrary functions such that $S \neq 0$
and $E \geq -1/2$ at all $r$.

The source in the Einstein equations is dust ($p = 0$) with the velocity field
$u^{\alpha} = {\delta_0}^{\alpha}$. The surfaces of constant $t$ and $r$ are
nonconcentric spheres, and $(x, y)$ are stereographic coordinates on each
sphere. At a fixed $r$, they are related to the spherical coordinates by
\begin{eqnarray}\label{2.3}
x &=& P + S \cot(\vartheta/2) \cos \varphi, \nonumber \\
y &=& Q + S \cot(\vartheta/2) \sin \varphi.
\end{eqnarray}
The functions $(P, Q, S)$ determine the centers of the spheres in the spaces of
constant $t$ (see illustrations in Ref. \cite{Kras2016b}). Because of the
nonconcentricity, the QSS spacetimes, in general, have no symmetry
\cite{BoST1977}.

With $\Lambda = 0$ assumed, $\Phi(t,r)$ obeys
\begin{equation}\label{2.4}
{\Phi,_t}^2 = 2 E(r) + \frac {2 M(r)} {\Phi},
\end{equation}
where $M(r)$ is an arbitrary function. We consider models with $E > 0$, then
\begin{eqnarray}\label{2.5}
\Phi(t,r) &=& \frac M {2E} (\cosh \eta - 1), \nonumber \\
\sinh \eta - \eta &=& \frac {(2E)^{3/2}} M \left[t - t_B(r)\right],
\end{eqnarray}
where $t_B(r)$ is one more arbitrary function; $t = t_B(r)$ is the BB time, at
which $\Phi(t_B, r) = 0$. We assume $\Phi,_t > 0$ (the Universe is expanding).

The mass density implied by (\ref{2.1}) is
\begin{equation}\label{2.6}
\kappa \rho = \frac {2 \left(M,_r - 3 M {\cal E},_r / {\cal E}\right)} {\Phi^2
\left(\Phi,_r - \Phi {\cal E},_r / {\cal E}\right)}, \quad \kappa \df \frac {8
\pi G} {c^2}.
\end{equation}
This density distribution is a mass dipole superposed on a spherically symmetric
monopole \cite{Szek1975b,DeSo1985}. The dipole, generated by ${\cal E},_r/{\cal
E}$, vanishes where ${\cal E},_r = 0$. The density is minimum where ${\cal
E},_r/{\cal E}$ is maximum and vice versa \cite{HeKr2002}.

The arbitrary functions must be such that $0 < \rho < \infty$ at all $t >
t_B(r)$. The conditions that ensure this are \cite{HeKr2002}:
\begin{eqnarray}
\frac {M,_r} {3M} &\geq& \frac {\sqrt{(S,_r)^2 + (P,_r)^2 + (Q,_r)^2}} S
~~~~\forall~r, \label{2.7} \\
\frac {E,_r} {2E} &>& \frac {\sqrt{(S,_r)^2 + (P,_r)^2 + (Q,_r)^2}} S
~~~~\forall~r. \label{2.8}
\end{eqnarray}
These inequalities imply \cite{HeKr2002}
\begin{equation}\label{2.9}
\frac {M,_r} {3M} \geq \frac {{\cal E},_r} {\cal E}, \qquad \frac {E,_r} {2E} >
\frac {{\cal E},_r} {\cal E} \qquad \forall~r.
\end{equation}
The extrema of ${\cal E},_r/{\cal E}$ with respect to $(x, y)$ are
\cite{HeKr2002}
\begin{equation}\label{2.10}
\left.\frac {{\cal E},_r} {\cal E}\right|_{\rm extreme} = \pm \frac
{\sqrt{(S,_r)^2 + (P,_r)^2 + (Q,_r)^2}} S,
\end{equation}
with $+$ corresponding to maximum and $-$ to minimum. In the following, we will
call these two loci ``dipole maximum'' and ``dipole minimum'', respectively.

The L--T models follow from the QSS models as the limit of constant $(P, Q, S)$.
Then the constant-$(t, r)$ spheres become concentric, and the spacetime becomes
spherically symmetric. The Friedmann limit is obtained when $E / M^{2/3}$ and
$t_B$ are constant (in this limit, $(P, Q, S)$ can be made constant by a
coordinate transformation). A QSS spacetime can be matched to a Friedmann
spacetime across an $r =$ constant hypersurface.

Because of $p = 0$, the QSS models can describe the past evolution of the
Universe no further back than to the last scattering hypersurface (LSH). See
Sec. \ref{corrM2} for information on how to determine it in our model.

\section{The QSS models considered in this paper}\label{QSShere}

We will consider such QSS spacetimes whose L--T limit is Model 2 of Ref.
\cite{Kras2016a}. The $r$-coordinate is chosen so that
\begin{equation}\label{3.1}
M = M_0 r^3,
\end{equation}
and $M_0 = 1$ (kept in formulae for dimensional clarity) \cite{Kras2014d}. From
this point on, the $r$-coordinate is unique. The function $E(r)$, assumed in the
form
\begin{equation}\label{3.2}
2E/r^2 \df -k = 0.4,
\end{equation}
is the same as in the background Friedmann model.

The units used in numerical calculations were introduced and justified in Ref.
\cite{Kras2014a}. Taking \cite{unitconver}
\begin{equation}\label{3.3}
1\ {\rm pc} = 3.086 \times 10^{13}\ {\rm km}, \quad 1\ {\rm y} = 3.156 \times
10^7\ {\rm s},
\end{equation}
the numerical length unit (NLU) and the numerical time unit (NTU) are defined as
follows:
\begin{equation}\label{3.4}
1\ {\rm NTU} = 1\ {\rm NLU} = 9.8 \times 10^{10}\ {\rm y} = 3 \times 10^4\ {\rm
Mpc}.
\end{equation}

\begin{figure}[h]
\begin{center}
\includegraphics[scale=0.5]{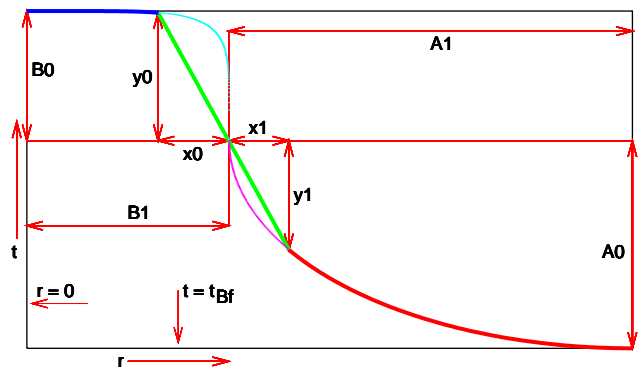}
\caption{Parameters of the bang-time profile in the quasispherical Szekeres
region; see text for explanation.}
 \label{drawpicture6}
\end{center}
\end{figure}

The BB profile belongs to the same 5-parameter family as in Ref.
\cite{Kras2016a}, see Fig. \ref{drawpicture6}. It consists of two curved arcs
and a straight line segment joining them. The upper-left arc, shown as a thicker
line, is a segment of the curve
\begin{equation}\label{3.5}
\frac {r^6} {{B_1}^6} + \frac {\left(t - t_{\rm Bf} - A_0\right)^6} {{B_0}^6} =
1,
\end{equation}
where
\begin{equation}\label{3.6}
t_{\rm Bf} = -0.13945554689046649\ {\rm NTU} \approx -13.67 \times 10^9\ {\rm
years};
\end{equation}
see Sec. \ref{backgrmodel} for comments on this value. The lower-right arc (also
shown as a thicker line) is a segment of the ellipse
\begin{equation}\label{3.7}
\frac {\left(r - B_1 - A_1\right)^2} {{A_1}^2} + \frac {\left(t - t_{\rm Bf} -
A_0\right)^2} {{A_0}^2} = 1.
\end{equation}
The straight segment\footnote{It was introduced to keep $\dril {t_B} r$ finite
everywhere.} passes through the point $(r, t) = (B_1, t_{\rm Bf} + A_0)$ where
the full curves (shown as dotted lines) would meet; $x_0$ determines its slope.

The free parameters are $A_0$, $A_1$, $B_0$, $B_1$ and $x_0$. Figure
\ref{drawpicture6} does not show the values used in numerical calculations; in
particular $x_0$ and $A_1$ are greatly exaggerated. The actual values in Model 2
of Ref. \cite{Kras2016a} are
\begin{equation} \label{3.8}
\left(\begin{array}{l}
A_0 \\
B_0 \\
A_1 \\
B_1 \\
x_0\\
\end{array}\right) = \left(\begin{array}{l}
 0.000026\ {\rm NTU} \\
 0.0001\ {\rm NTU} \\
 1 \times 10^{-10} \\
 0.015 \\
 2 \times 10^{-13} \\
 \end{array}\right)
\end{equation}
($A_1$, $B_1$ and $x_0$ are dimensionless). This profile will be the starting
point for modifications.

The QSS model used here is axially symmetric, with $P(r) = Q(r) = 0$ and $S(r)$
the same as in Ref. \cite{Kras2016b}:
\begin{equation}\label{3.9}
S = \sqrt{a^2 + r^2},
\end{equation}
where $a > 0$ is a constant, and so
\begin{equation}\label{3.10}
{\cal E} = \frac 1 {2 S}\ \left(x^2 + y^2 + S^2\right);
\end{equation}
This $S(r)$ obeys (\ref{2.7}) and (\ref{2.8}), which, using (\ref{3.1}) and
(\ref{3.2}), both reduce to
\begin{equation}\label{3.11}
1/r > S,_r / S.
\end{equation}
The equation of the dipole ``equator'' ${\cal E},_r = 0$ is
\begin{equation}\label{3.12}
x^2 + y^2 = S^2;
\end{equation}
the axis of symmetry is $x = y = 0$. The extrema of the dipole are, from
(\ref{2.10})
\begin{equation}\label{3.13}
\left.\frac {{\cal E},_r} {\cal E}\right|_{\rm extreme} = \pm \frac {S,_r} S.
\end{equation}

At $r > r_b$, where
\begin{equation}\label{3.14}
r_b = A_1 + B_1 = 0.0150000001,
\end{equation}
the BB profile becomes flat, and the geometry of the model becomes Friedmannian.
See Sec. \ref{symmetric} for remarks on the choice of coordinates in that
region.

\section{The background model}\label{backgrmodel}

\setcounter{equation}{0}

Our Friedmann background is defined by:
\begin{equation}\label{4.1}
\Lambda = 0, \qquad k = - 0.4, \qquad t_B = t_{\rm Bf},
\end{equation}
where $k$ is the curvature index and $t_B$ is the BB time given by (\ref{3.6});
$t = 0$ is the present time. The $t_{\rm Bf}$ is the asymptotic value of the
function $t_B(r)$ in the L--T model that mimicked accelerating expansion
\cite{Kras2014d}. This differs by $\sim 1.6 \%$ from $(- T)$, where $T$ is the
age of the Universe given by the Planck satellite team \cite{Plan2014}
\begin{equation}
T = 13.819 \times 10^9\ {\rm y} = 0.141\ {\rm NTU}. \label{4.2}
\end{equation}

The density at the last scattering time is \cite{Kras2016a}
\begin{equation}\label{4.3}
\kappa \rho_{\rm LS} = 56.1294161975316 \times 10^9 \ ({\rm NLU})^{-2}.
\end{equation}
This value follows from the model of the cosmological recombination process
\cite{Peeb1968,ZKSu1968,recoWiki} and is independent of the after-recombination
model. With (\ref{4.1}), $\rho_{\rm LS}$ implies the redshift relative to the
present time
\begin{equation}\label{4.4}
1 + z^{\rm b}_{\rm LS} = 952.611615159.
\end{equation}
This differs by $\sim 12.7 \%$ from the $\Lambda$CDM value
\cite{Plan2014,Plan2014b}
\begin{equation}\label{4.5}
z_{\rm LS} = 1090.
\end{equation}
The present temperature of the CMB radiation is directly measured, so if
(\ref{4.4}) were taken for real, the temperature of the background radiation at
emission would be $\sim 3380$ K instead of $\sim 3000$ K dictated by current
knowledge. To reconcile our model with these data, many recalculations would be
required. Since our model needs other improvements anyway, we will stick to
(\ref{4.1}), to be able to compare the present results with the earlier ones.

\section{Null geodesics in the axially symmetric QSS
spacetimes}\label{symmetric}

\setcounter{equation}{0}

In an axially symmetric QSS metric, $x$ and $y$ can be chosen such that $P = Q =
0$; then $x = y = 0$ is the symmetry axis \cite{BKHC2010,NoDe2007}. However, the
loci $x = \infty$ and $y = \infty$ are coordinate singularities (they are at the
pole of the stereographic projection), and numerical integration of nonaxial
geodesics breaks down on crossing those sets. Therefore, we introduce the new
coordinates $(\vartheta, \varphi)$ by
\begin{equation}\label{5.1}
x = S_b \cot(\vartheta/2) \cos \varphi, \qquad y = S_b \cot(\vartheta/2) \sin
\varphi,
\end{equation}
where $S_b$ is $S$ at the Szekeres/Friedmann boundary:
\begin{equation}\label{5.2}
S_b \df S(r_b) = \sqrt{a^2 + {r_b}^2}.
\end{equation}
This changes (\ref{2.1}) and (\ref{2.2}) to
\begin{equation}\label{5.3}
{\rm d} s^2 = {\rm d} t^2 - \frac {{\cal N}^2 {\rm d} r^2} {1 + 2 E(r)} -
\left(\frac {\Phi} {\cal F}\right)^2 \left({\rm d} \vartheta^2 + \sin^2
\vartheta {\rm d} \varphi^2\right),
\\
\end{equation}
\begin{eqnarray}
{\cal F} &\df& \frac {S_b} {2 S}\ (1 + \cos \vartheta) + \frac S {2 S_b}\ (1 -
\cos \vartheta), \nonumber \\
{\cal N} &\df& \Phi,_r - \Phi {\cal F},_r/{\cal F}. \label{5.4}
\end{eqnarray}
The dipole equator ${\cal F},_r = 0$ is now at $\cot (\vartheta_{\rm eq}/2) =
S/S_b$ (so $\vartheta_{\rm eq} = \pi/2$ at the QSS boundary). On the boundary
sphere $r = r_b$ we have ${\cal F} = 1$ and $(\vartheta, \varphi)$ become the
spherical coordinates with the origin at $r = 0$.

Along a geodesic we denote
\begin{equation}\label{5.5}
\left(k^t, k^r, k^{\vartheta}, k^{\varphi}\right) \df \dr {(t, r, \vartheta,
\varphi)} {\lambda},
\end{equation}
where $\lambda$ is an affine parameter. The geodesic equations for (\ref{5.3})
-- (\ref{5.4}) are
\begin{equation}\label{5.6}
\dr {k^t} {\lambda} + \frac {{\cal N} {\cal N},_t} {1 + 2E} \left(k^r \right)^2
+ \frac{\Phi {\Phi,_t}}{{\cal F}^2} \left[\left(k^{\vartheta}\right)^2 + \sin^2
\vartheta \left(k^{\varphi}\right)^2\right] = 0,
\end{equation}
\begin{eqnarray}
\dr {k^r} {\lambda} &+& 2 \frac {{\cal N},_t} {\cal N} k^t k^r \nonumber \\
&+& \left(\frac {{\cal N},_r} {\cal N} - \frac {E,_r} {1 + 2E}\right)
\left(k^r\right)^2 + 2 \frac {S,_r \sin \vartheta \Phi} {S {\cal F}^2 {\cal N}}\
k^r k^{\vartheta} \nonumber \\
&-& \frac {\Phi (1 + 2E)} {{\cal F}^2 {\cal N}}
\left[\left(k^{\vartheta}\right)^2 + \sin^2 \vartheta
\left(k^{\varphi}\right)^2\right] = 0, \label{5.7} \\
\dr {k^{\vartheta}} {\lambda} &+& 2 \frac {\Phi,_t} {\Phi} k^t k^{\vartheta} -
\frac {S,_r \sin \vartheta {\cal N}} {S \Phi (1 + 2E)}\ \left(k^r\right)^2 + 2
\frac {\cal N} {\Phi} k^r k^{\vartheta} \nonumber \\
&+& \frac {{\cal F},_{\vartheta}} {\cal F}\ \left[- \left(k^{\vartheta}\right)^2
+ \sin^2 \vartheta \left(k^{\varphi}\right)^2\right] \nonumber \\
&-& \cos \vartheta \sin \vartheta \left(k^{\varphi}\right)^2 = 0, \label{5.8} \\
\dr {k^{\varphi}} {\lambda} &+& 2 \frac {\Phi,_t} {\Phi} k^t k^{\varphi} + 2
\frac {\cal N} {\Phi} k^r k^{\varphi} \nonumber \\
&+& 2 \left[\frac {\cos \vartheta} {\sin \vartheta} - \frac {{\cal
F},_{\vartheta}} {\cal F}\right] k^{\vartheta} k^{\varphi} = 0. \label{5.9}
\end{eqnarray}
The geodesics determined by (\ref{5.6}) -- (\ref{5.9}) are null when
\begin{equation}\label{5.10}
\left(k^t\right)^2 - \frac {{\cal N}^2 \left(k^r\right)^2} {1 + 2E(r)} -
\left(\frac {\Phi} {\cal F}\right)^2 \left[\left(k^{\vartheta}\right)^2 + \sin^2
\vartheta \left(k^{\varphi}\right)^2\right] = 0.
\end{equation}
Note that $k^{\varphi} \equiv 0$ is a solution of (\ref{5.9}) while $\vartheta
\equiv 0$ and $\vartheta \equiv \pi$ (axial rays) are solutions of (\ref{5.8}).

To calculate $k^r$ on nonaxial null geodesics, Eq. (\ref{5.10}) will be used,
which is insensitive to the sign of $k^r$. A numerical program for integrating
the set \{(\ref{5.6}), (\ref{5.8}) -- (\ref{5.10})\} will have to change the
sign of $k^r$ wherever $k^r$ reaches zero.

There exist no null geodesics on which $k^{\varphi} \equiv 0$ and $\vartheta$
has any constant value other than 0 or $\pi$. This follows from (\ref{5.8}):
Suppose $k^{\varphi} \equiv 0$ everywhere and $k^{\vartheta} = 0$ at a point.
Then, if $\sin \vartheta \neq 0$, the third term in (\ref{5.8}) will be nonzero
(because $\left|S \Phi (1 + 2E)\right| < \infty$, $S,_r \neq 0$ from
(\ref{3.9}), ${\cal N} \neq 0$ from no-shell-crossing conditions \cite{HeKr2002}
and $k^r \neq 0$ from (\ref{5.10})), and so $\dril {k^{\vartheta}} {\lambda}
\neq 0$. Consequently, in the axially symmetric case the only analogues of
radial directions are $\vartheta = 0$ and $\vartheta = \pi$. The fact reported
under (\ref{6.4}) below is consistent with this.

The coefficient $1/\Phi$ in (\ref{5.8}) and (\ref{5.9}) becomes infinite at $r =
0$, where $\Phi = 0$ \cite{Kras2016a}, but all the suspicious-looking terms are
in fact finite there \cite{Kras2016b}. In the present paper the only geodesics
running through $r = 0$ will be the axial ones, on which (\ref{5.8}) and
(\ref{5.9}) are obeyed identically.

Let the subscript $o$ refer to the observation point. On past-directed rays $k^t
< 0$ and the affine parameter along each one can be chosen such that
\begin{equation}\label{5.11}
k^t_o = -1.
\end{equation}
Then, from (\ref{5.10}) we have
\begin{equation}\label{5.12}
\left(k_o^{\vartheta}\right)^2 + \sin^2 \vartheta \left(k_o^{\varphi}\right)^2
\leq \left(\frac {{\cal F}_o} {\Phi_o}\right)^2;
\end{equation}
the equality occurs when the ray is tangent to a hypersurface of constant $r$ at
the observation event, $k_o^r = 0$.

On the boundary $r = r_b$ between the QSS and Friedmann regions the coordinates
on both sides must coincide. Thus, for the Friedmann region one must use the
metric (\ref{5.3}) with $t_B = t_{\rm Bf}$ given by (\ref{3.6}) ($E$ has the
Friedmann form (\ref{3.2}) everywhere). The metric then becomes Friedmann with
no further limitation on $S$. But for correspondence with Ref. \cite{Kras2016a},
we choose the coordinates in the Friedmann region so that
\begin{equation}\label{5.13}
S = \sqrt{a^2 + {r_b}^2} = S_b.
\end{equation}
Then, ${\cal F} = 1$ and $(\vartheta, \varphi)$ are the spherical coordinates
throughout the Friedmann region.

\section{The redshift in axially symmetric QSS spacetimes}\label{axreds}

\setcounter{equation}{0}

Along a ray emitted at $P_e$ and observed at $P_o$
\begin{equation}\label{6.1}
1 + z = \frac {\left(u_{\alpha} k^{\alpha}\right)_e} {\left(u_{\alpha}
k^{\alpha}\right)_o},
\end{equation}
where $u_{\alpha}$ are the four-velocities of the emitter and of the observer,
and $k^{\alpha}$ is the affinely parametrised tangent vector field to the ray
\cite{Elli1971}. In our case, both $u_{\alpha} = {\delta^0}_{\alpha}$, and then
(\ref{6.1}) simplifies to $1 + z = {k_e}^t/{k_o}^t$. If the affine parameter is
rescaled so that (\ref{5.11}) holds, then
\begin{equation}\label{6.2}
1 + z = - {k_e}^t.
\end{equation}
Equation (\ref{5.9}) has the first integral:
\begin{equation}\label{6.3}
k^{\varphi} \sin^2 \vartheta \Phi^2 / {\cal F}^2 = J_0,
\end{equation}
where $J_0$ is constant. When (\ref{6.3}) is substituted in (\ref{5.10}), the
following results:
\begin{equation}\label{6.4}
(k^t)^2 = \frac {{\cal N}^2 \left(k^r\right)^2} {1 + 2E} + \left(\frac {\Phi}
{\cal F}\right)^2 \left(k^{\vartheta}\right)^2 + \left(\frac {J_0 {\cal F}}
{\sin \vartheta \Phi}\right)^2.
\end{equation}
Equations (\ref{6.4}) and (\ref{6.2}) show that for rays emitted at the BB,
where $\Phi = 0$, the observed redshift is infinite when $J_0 \neq 0$. A
necessary condition for infinite blueshift ($1 + z_o = 0$) is thus $J_0 = 0$, so

(a) either $k^{\varphi} = 0$, i.e. the ray proceeds in the hypersurface of
constant $\varphi$,

(b) or $\vartheta = 0, \pi$ along the ray ($J_0/\sin \vartheta \to 0$ when
$\vartheta \to 0, \pi$ by (\ref{6.3})).

\noindent Condition (b) appears to be also sufficient, but this has been
demonstrated only numerically in concrete examples of QSS models
(\cite{Kras2016b} and Sec. \ref{corrM2} here).

Consider a ray proceeding from event $P_1$ to $P_2$ and then from $P_2$ to
$P_3$. Denote the redshifts acquired in the intervals $[P_1, P_2]$, $[P_2, P_3]$
and $[P_1, P_3] = [P_1, P_2] \cup [P_2, P_3]$ by $z_{12}$, $z_{23}$ and
$z_{13}$, respectively. Then, from (\ref{6.1})
\begin{equation} \label{6.5}
1 + z_{13} = \left(1 + z_{12}\right) \left(1 + z_{23}\right).
\end{equation}
In particular, for a ray proceeding to the past from $P_1$ to $P_2$, and then
back to the future from $P_2$ to $P_1$:
\begin{equation}\label{6.6}
1 + z_{12} = \frac 1 {1 + z_{21}}.
\end{equation}

\section{The Extremum Redshift Surface}\label{ERS}

\setcounter{equation}{0}

Consider a null geodesic that stays in the surface $\{\vartheta, \varphi\} =
\{\pi, {\rm constant}\}$; it obeys (\ref{5.8}) and (\ref{5.9}) identically. On
it, $k^r \neq 0$ at all points because with $k^{\vartheta} = k^{\varphi} = 0$
the geodesic would be timelike wherever $k^r = 0$, so $r$ can be used as a
parameter. Assume the geodesic is past-directed so that (\ref{6.2}) applies.
Using (\ref{6.2}) and changing the parameter to $r$, we obtain from (\ref{5.6})
\begin{equation}\label{7.1}
\dr z r = \frac {{\cal N} {\cal N},_t} {1 + 2E}\ k^r.
\end{equation}
Since ${\cal N} \neq 0$ from no-shell-crossing conditions \cite{HeKr2002} and
$k^r \neq 0$, the extrema of $z$ on such a geodesic occur where
\begin{equation}\label{7.2}
{\cal N},_t \equiv \Phi,_{tr} - \Phi,_t {\cal F},_r/{\cal F} = 0.
\end{equation}
In deriving (\ref{7.2}), $\vartheta = \pi$ was assumed, but $\varphi$ was an
arbitrary constant. Thus, the set in spacetime defined by (\ref{7.2}) is
2-dimensional; it is the Extremum Redshift Surface (ERS) \cite{Kras2016b}.

{}From (\ref{2.4}) and (\ref{3.2}) we obtain
\begin{eqnarray}
\Phi,_t &=& r \sqrt{\frac {2 M_0 r} {\Phi} - k}, \label{7.3}\\
\Phi,_{tr} &=& \sqrt{\frac {2 M_0 r} {\Phi} - k} + \frac {M_0 r^3} {\Phi^2}\
t_{B,r}. \label{7.4}
\end{eqnarray}
Using (\ref{7.3}), (\ref{7.4}) and (\ref{5.4}) with $\vartheta = \pi$, Eq.
(\ref{7.2}) becomes
\begin{equation}\label{7.5}
\sqrt{\frac {2 M_0 r} {\Phi} - k} \left(1 - r \frac {S,_r} S\right) = - \frac
{M_0 r^3} {\Phi^2}\ t_{B,r}.
\end{equation}
To avoid shell crossings, $t_{B,r} < 0$ must hold at all $r > 0$
\cite{HeKr2002}, \cite{PlKr2006},\footnote{Refs. \cite{HeKr2002} and
\cite{PlKr2006} did not spell out the condition $r > 0$ in deriving the
no-shell-crossing conditions, but it is implicitly there.} so the right-hand
side of (\ref{7.5}) is non-negative. The left-hand side is positive with $S$
given by (\ref{3.9}). Using (\ref{2.5}) for $\Phi$, remembering that $k < 0$ and
denoting
\begin{equation}\label{7.6}
\chi \df \sinh^2 (\eta/2)
\end{equation}
we obtain from (\ref{7.5})
\begin{equation}\label{7.7}
\chi^4 + \chi^3 = - k^3 \left[\frac {r t_{B,r}} {4 M_0 \left(1 - r S,_r /
S\right)}\right]^2.
\end{equation}
With $k < 0$, (\ref{7.7}) is solvable for $\chi$ at any $r$, since its left-hand
side is independent of $r$ and can vary from 0 to $+\infty$ while the right-hand
side is non-negative.

Note that where $t_{B,r} = 0$, Eqs. (\ref{7.7}) and (\ref{7.6}) imply $\chi =
\eta = 0$, i.e. at those points the ERS is tangent to the BB. Also, the ERS is
tangent to the BB at $r = 0$ unless $\dril {t_B} r \llim{r \to 0} \infty$. (This
would imply $\dril {\rho} r \llim{r \to 0} \infty$, an infinitely thin peak in
density at $r = 0$ -- an unusual configuration, but {\em not a curvature
singularity} \cite{KHBC2010}.) The model considered here will have $t_{B,r} = 0$
at $r = 0$.

In the limit $S,_r = 0$, (\ref{7.7}) reproduces the equation of the Extremum
Redshift {\em Hyper}surface (ERH) of Ref. \cite{Kras2014d}.

Equation (\ref{7.7}) was derived for null geodesics proceeding along $\vartheta
= \pi$, where ${\cal F},_r/{\cal F} = S,_r/S > 0$. With $S$ given by (\ref{3.9})
we have
\begin{equation}\label{7.8}
F_1 \df 1/\left(1 - rS,_r/S\right) = (r/a)^2 + 1 > 1,
\end{equation}
so, at a given $r$, the ERS has a greater $\eta$ (and so a greater $t - t_B$)
than the corresponding ERH of the L--T model. Also, the extrema of $z$ along the
dipole maximum occur at a greater $\chi$ (and thus greater $t - t_B$) when $a$
is smaller. This will be illustrated by Fig. \ref{smallz} in the next section.

Conversely, for a ray proceeding along the dipole minimum axis (where $\vartheta
= 0$), the factor $F_1$ is replaced by
\begin{equation}\label{7.9}
F_2 \df 1/\left(1 + rS,_r/S\right) = \frac {a^2 + r^2} {a^2 + 2r^2} < 1,
\end{equation}
and so the ERS has a {\em smaller} $t - t_B$ than the ERH in L--T. Also here, a
smaller $a$ has a more pronounced effect.

Extrema of redshift also exist along directions other than $\vartheta = 0$ and
$\vartheta = \pi$, as will be demonstrated by numerical examples in Sec.
\ref{noaxreds}, but a general equation defining their loci remains to be
derived.

\section{A generalised Model 2 of Ref. \cite{Kras2016a}}\label{corrM2}

\setcounter{equation}{0}

Along each past-directed null geodesic, the mass density is calculated using
(\ref{2.5}) -- (\ref{2.6}). As explained in Sec. \ref{backgrmodel}, in any model
the density at the LSH must be the same as in (\ref{4.3}). So, the instant of
crossing the LSH is that where the density becomes equal to (\ref{4.3}).

The starting point for this paper is Model 2 of Ref. \cite{Kras2016a}, whose
functions $M(r)$, $E(r)$ and $t_B(r)$ are given by (\ref{3.1}), (\ref{3.2}) and
(\ref{3.5}) -- (\ref{3.8}). In that model, the strongest blueshift between the
LSH and the present epoch was
\begin{equation}\label{8.1}
1 + z_{\rm maxb} = 1.36167578 \times 10^{-5}.
\end{equation}
It was calculated by the rule (\ref{6.5}). The first factor,
\begin{equation}\label{8.2}
1 + z_{\rm ols2} = 1.07858890707746014 \times 10^{-7},
\end{equation}
was the blueshift between the LSH and $r = 0$, achieved on a path that will be
called ``Ray A''. The second factor,
\begin{equation}\label{8.3}
1 + z_{\rm po2} = 126.246039921.
\end{equation}
was the redshift between $r = 0$ and the present epoch on a path going off from
the same initial point as Ray A, but to the future; it will be called ``Ray B''.

On Model 2, axially symmetric QSS deformations given by $P = Q = 0$, (\ref{3.9})
and (\ref{3.10}) are superposed. Numerical experiments with rays proceeding
along $\vartheta = \pi$ were done to improve on (\ref{8.1}) as much as possible.
As explained under (\ref{7.8}), smaller $a$ increases the region under the ERS.
So, with the parameters of (\ref{3.8}), $a^2$ was gradually changed from 10
through 1, 1/10, $10^{-2}$, $10^{-3}$ to $10^{-4}$. For each $a$ the quantity
\begin{equation}\label{8.4}
t(0) - t_B(0) \df \Delta t_c
\end{equation}
was chosen such as to obtain a minimum $1 + z$ between the LSH and $r = 0$. This
led to smaller $1 + z$ on Ray A only down to $a^2 = 0.001$. With $a^2$ still
smaller, the ray either flew over the BB hump and crossed the LSH in the
Friedmann region with a large $z > 0$ or dipped under the LSH still within the
QSS region with a small $z > 0$. No intermediate value of $\Delta t_c$ led to $z
< 0$ (but this discontinuity could possibly be overcome with greater numerical
precision). The best result achieved with $a^2 = 0.001$ was $1 + z_2 =
8.87933914173189009 \times 10^{-8}$.

In the next experiments, the slope of the straight segment of the BB profile was
gradually decreased, i.e $x_0$ was increased from $2 \times 10^{-13}$ through $1
\times 10^{-12}$ to $1 \times 10^{-11}$, with the other parameters unchanged.
For each value of $x_0$, the $\Delta t_c$ leading to the smallest $1 + z$ was
determined. The best result achieved at this stage was
\begin{equation}\label{8.5}
1 + z_1 = 6.74014204449235876 \times 10^{-8}.
\end{equation}
Varying $A_1$, $B_1$, $B_0$, and lowering the degree of (\ref{3.5}) to 4 and to
2, led to nothing better than (\ref{8.5}). So, this is taken as the best
improvement over the L--T model achieved using an axially symmetric QSS
deformation.

\begin{figure}[h]
\begin{center}
 \includegraphics[scale=0.5]{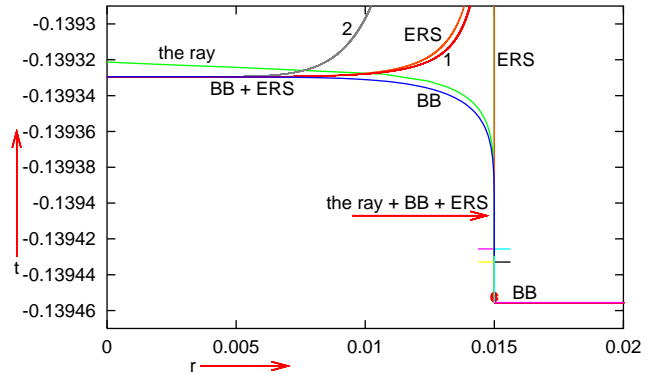}
\caption{The Big Bang profile and the axial ray with the smallest $1 + z$ in the
Szekeres model. The short horizontal strokes are at the ends of the straight BB
segment. The dot shows where the ray hits the LSH. See text for more
explanation.}
 \label{smallz}
\end{center}
\end{figure}

Figure \ref{smallz} shows Ray A, with $1 + z_1$ given by (\ref{8.5}), and the
corresponding ERS and BB profiles. Curve 1 is the ERH profile of Model 2 from
Ref. \cite{Kras2016a}, and Curve 2 is the ERS profile with $a^2 = 10^{-5}$. As
stated above, smaller $a$ gives more space under the ERS, but when too small it
creates a discontinuity in $z$ that prevents $z < 0$ altogether.

The ERS profile has two branches on each side of $r = 0$, so some rays will
intersect it four times and $z$ along them will have two local maxima and two
local minima. Examples will appear in Sec. \ref{noaxreds}.

On Ray B, the upward $1 + z$ is
\begin{equation}\label{8.6}
1 + z_{\rm 1\ up} = 6.39228356761256666 \times 10^{-3}.
\end{equation}
Thus, total $(1 + z)$ between LSH and now is
\begin{equation}\label{8.7}
1 + z_2 = \frac {1 + z_1} {1 + z_{\rm 1\ up}} = 1.05441849899 \times 10^{-5}\ \
\ \ .
\end{equation}
This fits the lowest-frequency GRBs, for which \cite{Kras2016a}
\begin{equation}\label{8.8}
1 + z_{\rm max} \approx 1.689 \times 10^{-5},
\end{equation}
with a wider margin than (\ref{8.1}), so the BB hump can now be lowered to yield
$(1 + z)$ closer to (\ref{8.8}). The easiest way to do this is to decrease $B_0$
(see Fig. \ref{drawpicture6}). Then $\Delta t_c$ is fine-tuned to make $(1 +
z_{\rm ols})$ on Ray A as small as possible ($1 + z_{\rm ols}$ gets larger when
$B_0$ gets smaller, so there is a limit on decreasing $B_0$). The $B_0$ that
allows sufficiently small $(1 + z)_{\rm ols}$ is
\begin{equation}\label{8.9}
B_0 = 0.000091,
\end{equation}
and then the smallest $1 + z$ on Ray A is
\begin{equation}\label{8.10}
1 + z_{\rm ols3} = 1.11939135405414447 \times 10^{-7}.
\end{equation}

For Ray B corresponding to Ray A of (\ref{8.10}) (proceeding along the dipole
minimum), the $1 + z$ between $r = 0$ and the present epoch is
\begin{equation}\label{8.11}
1 + z_{\rm 3\ up} = 7.11151887923544557 \times 10^{-3},
\end{equation}
so $(1 + z)$ between LSH and now along Rays A and B is
\begin{equation}\label{8.12}
1 + z_3 = \frac {1 + z_{\rm ols3}} {1 + z_{\rm 3\ up}} = 1.574 \times 10^{-5},
\end{equation}
and the present observer is at
\begin{equation}\label{8.13}
r = r_{\rm obs} = 0.88983013520392229.
\end{equation}
This is larger than $r_{\rm O2} = 0.88705643159726955$ in Model 2 of Ref.
\cite{Kras2016a}. Thus, a Szekeres deformation superposed on an L--T model
results in moving the observer further from the radiation source, which leads to
a smaller angular diameter of the source seen in the sky; see Sec. \ref{noax}.

Rays A and B referred to above have
\begin{equation}\label{8.14}
\Delta t_c = 0.00000863099500\ {\rm NTU}.
\end{equation}

The corresponding results for rays propagating in the opposite direction, i.e.
along the dipole minimum between the LSH and $r = 0$ (Ray C), and along the
dipole maximum between $r = 0$ and the observer (Ray D), are as follows. The
best value of $1 + z$ on Ray C is
\begin{equation}\label{8.15}
1 + z_{\rm 1\ dip\ min} = 1.73185662921682137 \times 10^{-7},
\end{equation}
achieved with
\begin{equation}\label{8.16}
\Delta t_c = 0.00000981550000\ {\rm NTU}.
\end{equation}
Then, $1 + z$ calculated toward the future along the dipole maximum is
\begin{equation}\label{8.17}
1 + z_{\rm 2\ dmax\ up} = 7.26948511585012724 \times 10^{-3}.
\end{equation}
So, the $1 + z$ between the LSH and the present time is
\begin{equation}\label{8.18}
1 + z_4 = \frac {1 + z_{\rm 1\ dip\ min}} {1 + z_{\rm 2\ dmax\ up}} = 2.382
\times 10^{-5}.
\end{equation}
The present time was reached by the ray at
\begin{equation}\label{8.19}
\widetilde{r}_{\rm obs} = 0.88935629118490100.
\end{equation}
Thus, on this ray $1 + z$ is larger while $r_{\rm obs}$ is smaller.

In each case the numerical calculation overshot the present time. For the ray
that produced (\ref{8.11}) and (\ref{8.13}), the value of $t$ at the endpoint
was
\begin{equation}\label{8.20}
t_{\rm end\ 1} = 5.75302117391131287 \times 10^{-11}\ {\rm NTU},
\end{equation}
and for the ray that produced (\ref{8.17}) and (\ref{8.19}) it was
\begin{equation}\label{8.21}
t_{\rm end\ 2} = 9.65282969667925857 \times 10^{-10}\ {\rm NTU}.
\end{equation}

\section{Nonaxial plane rays}\label{noax}

\setcounter{equation}{0}

So far, rays crossing the symmetry axis of the $t =$ constant spaces in the
metric (\ref{5.3}) -- (\ref{5.4}) were considered. Now, we will consider
nonaxial rays ($\vartheta$ will no longer be 0 or $\pi$ all along the ray)
propagating in a hypersurface of constant $\varphi$. By (\ref{6.3}), $J_0 = 0$
along them, and they obey (\ref{5.9}) identically. Because of axial symmetry of
the model, the image will be the same for every $\varphi$.

We will consider pencils of rays flying through the vicinity of the BB hump
shown in Fig. \ref{smallz} and reaching the present observer situated in three
locations:

Observer I: At
\begin{equation}\label{9.1}
(t, r, \vartheta)_{\rm I} = (t_{\rm end\ 1}, r_{\rm obs}, 0),
\end{equation}
with $r_{\rm obs}$ given by (\ref{8.13}). This is the endpoint of Ray B.

Observer II: At
\begin{equation}\label{9.2}
(t, r, \vartheta)_{\rm II} = (t_{\rm end\ 2}, \widetilde{r}_{\rm obs}, \pi),
\end{equation}
with $\widetilde{r}_{\rm obs}$ given by (\ref{8.19}). This is the endpoint of
Ray D.

Observer III: At
\begin{equation}\label{9.3}
(t, r, \vartheta)_{\rm III} = (0, r_p, \pi/2), \quad {\rm where} \quad r_p =
(r_{\rm obs} + \widetilde{r}_{\rm obs})/2.
\end{equation}
The $\vartheta_{\rm III}$ is at the dipole equator on the boundary of the
Szekeres region. One ray reaching Observer III will have $\vartheta = \pi/2$
throughout the Friedmann region.

The equations to be integrated are, from (\ref{5.6}) -- (\ref{5.10}):
\begin{eqnarray}
\dr t {\lambda} &=& k^t, \label{9.4} \\
\dr {k^t} {\lambda} &=& - \frac {{\cal N} {\cal N},_t} {1 + 2E} \left(k^r
\right)^2 - \frac{\Phi {\Phi,_t}}{{\cal F}^2} \left(k^{\vartheta}\right)^2,
\label{9.5} \\
\dr {\vartheta} {\lambda} &=& k^{\vartheta}, \label{9.6} \\
\dr {k^{\vartheta}} {\lambda} &=& - 2 \frac {\Phi,_t} {\Phi} k^t k^{\vartheta} +
\frac {\sin \vartheta S,_r {\cal N}} {S \Phi (1 + 2E)}\ \left(k^r\right)^2 - 2
\frac {\cal N} {\Phi} k^r k^{\vartheta}  \nonumber \\
&+& \frac {\sin \vartheta \left(S^2 - {S_b}^2\right)} {2S S_b {\cal F}}\
\left(k^{\vartheta}\right)^2, \label{9.7} \\
\dr r {\lambda} &=& k^r, \label{9.8} \\
k^r &=& \pm \frac {\sqrt{1 + 2E}} {\cal N} \sqrt{\xi}, \nonumber \\
\xi &\df& \left(k^t\right)^2 - \left(\frac {\Phi k^{\vartheta}} {\cal
F}\right)^2. \label{9.9}
\end{eqnarray}
The initial values for $(t, r, \vartheta)$ will be at the observer positions
specified above, the initial value for $k^t$ is (\ref{5.11}), and the rays will
be calculated backward in time from there. With $k^{\varphi} = 0$, Eq.
(\ref{5.12}) reduces to
\begin{equation}\label{9.10}
\left(k_o^{\vartheta}\right)^2 \leq \left(\frac {{\cal F}_o} {\Phi_o}\right)^2.
\end{equation}
As before, the equality occurs when $k^r_o = 0$.

For observers in the Friedmann region, ${\cal F}_o = 1$, as explained under Eq.
(\ref{5.13}). For Observer I $\Phi_o$ was calculated by the program that found
(\ref{8.11}); it is
\begin{equation}\label{9.11}
(\Phi_o)_{\rm obs\ 1} = 0.40202832540890049.
\end{equation}

The angle $\alpha$ between two rays at an observer can be calculated as follows.
The direction of a ray is determined by the unit spacelike vector given by
\cite{PlKr2006}
\begin{equation}\label{9.12}
n^{\alpha} = u^{\alpha} - \frac {k^{\alpha}} {k^{\rho} u_{\rho}},
\end{equation}
where $k^{\alpha}$ is the tangent vector to the ray and $u^{\alpha}$ is the
velocity vector of the observer; $n^{\alpha} u_{\alpha} = 0$. Since $g_{\alpha
\beta} n^{\alpha} n^{\beta} = -1$, the angle between two directions obeys
\begin{equation}\label{9.13}
\cos \alpha = - g_{\alpha \beta} n^{\alpha}_1 n^{\beta}_2.
\end{equation}
Since $u^{\alpha} = {\delta^{\alpha}}_0$ everywhere, and $k^0 = -1$ at the
observer, the components of a general $n^{\alpha}$ at the observer are
\begin{equation}\label{9.14}
n^{\alpha}_o = \left(0, k^r_o, k^{\vartheta}_o, k^{\varphi}_o\right).
\end{equation}
Using (\ref{9.9}), (\ref{5.3}) and assuming $k^{\varphi}_o = 0$ we then obtain
for the angle $\alpha_{RS}$ between rays R and S
\begin{eqnarray}\label{9.15}
\cos \alpha_{RS} &=& \sqrt{1 - \left(\frac {k^{\vartheta}_{Ro} \Phi_o} {{\cal
F}_o}\right)^2} \times \sqrt{1 - \left(\frac {k^{\vartheta}_{So} \Phi_o} {{\cal
F}_o}\right)^2} \nonumber \\
&+& k^{\vartheta}_{Ro} k^{\vartheta}_{So} \left(\frac {\Phi_o} {{\cal
F}_o}\right)^2.
\end{eqnarray}
Both $k^{\vartheta}_o$ must obey (\ref{9.10}), so $\left|\cos \alpha_{RS}\right|
\leq 1$ and $\alpha_{RS}$ obeying (\ref{9.15}) exists.

When Ray R is axial ($k^{\vartheta}_{Ro} = 0$), and the observer lies in the
Friedmann region where ${\cal F}_o = 1$, (\ref{9.15}) becomes
\begin{equation}\label{9.16}
\cos \alpha_{RS} = \sqrt{1 - \left(k^{\vartheta}_{So} \Phi_o\right)^2} \quad
\Longrightarrow \quad \sin \alpha_{RS} = k^{\vartheta}_{So} \Phi_o.
\end{equation}
This equation can be used to estimate the angular radius of a radiation source
in the sky; then $\alpha$ is the angle between the direction of the central ray
(going along the symmetry axis for Observers I and II) and the direction of the
ray that grazes the edge of the source. The latter can be approximately
determined in numerical experiments.

The redshift in the Friedmann background between the LSH and the present time,
calculated numerically along a null geodesic is
\begin{equation}\label{9.17}
1 + z_b = 951.83531161489873.
\end{equation}
This differs slightly from (\ref{4.4}), which was calculated from $1 + z^{\rm
b}_{\rm LS} = {\cal R}_{\rm now}/{\cal R}_{\rm LS}$, where ${\cal R}$ is the
Friedmann scale factor, and also from $1 + z_{\rm comp} = 951.91469714961829$
calculated in Ref. \cite{Kras2016a}. The differences are caused by numerical
inaccuracies (in particular, a different numerical step was used in
\cite{Kras2016a}). Since all null geodesics in the following will be calculated
numerically, (\ref{9.17}) will be taken as the reference value.

The figures in this section show rays that stay over or near the BB hump for
some of the flight time. The initial value of $k^r$ for each ray follows from
(\ref{9.9}) after the value of $k_o^{\vartheta}$ is chosen. At all initial
points, $k^r < 0$, but $\xi$ was monitored along each ray, and when it went down
to or below zero, the sign of $k^r$ was reversed.\footnote{Note that $\xi < 0$
is impossible on a null geodesic with $k^{\varphi} = 0$ by (\ref{5.10}). But it
can happen because of numerical inaccuracy. If $\xi < 0$ at step $n$, then for
this step it was replaced by $(- \xi)$; then it should begin to grow. Along some
rays the sign reversals of $\xi$ in a vicinity of the smallest $r$ had to be
done many times .}

\subsection{Rays reaching Observer I}

Table \ref{obsIrays} lists the parameters of exemplary nonradial rays received
by Observer I, with the angular radii calculated by (\ref{9.16}). The angular
radius of the whole BB hump (Ray 9 in the table) here is somewhat smaller than
the $1.00097^{\circ}$ in the L--T/Friedmann model of Ref. \cite{Kras2016a}.
Decreasing this radius was one of the aims of replacing the L--T region with
Szekeres.

\begin{table}[h]
\begin{center}
\caption{Parameters of nonaxial rays reaching Observer I. For Ray 9,
$k^{\vartheta}_o = 0.042007485$}
\bigskip
\begin{tabular}{|c|c|c|c|}
  \hline \hline
  Ray & $k^{\vartheta}_o$ & Angular radius ($^{\circ}$) & $1 + z$ at LSH \\
  \hline \hline
  0 & 0.000001 & $2.3 \times 10^{-5}$ & 294.74391009044683 \\
  \hline
  1 & 0.0005 & 0.0115 & 296.54474209835132 \\
  \hline
  2 & 0.002 & 0.046 & 304.52122850647874 \\
  \hline
  3 & 0.005 & 0.115 & 372.37434100449173 \\
  \hline
  4 & 0.009 & 0.207 & 541.61077498481632 \\
  \hline
  5 & 0.012 & 0.276 & 693.38900192388246 \\
  \hline
  6 & 0.02 & 0.461 & 906.63699789072280 \\
  \hline
  7 & 0.03 & 0.691 & 971.70020743827149 \\
  \hline
  8 & 0.035 & 0.806 & 981.87561752691374 \\
  \hline
  9 & 0.042 & 0.96767 & 951.83290067586029 \\
  \hline \hline
\end{tabular} \\
 \label{obsIrays}
\end{center}
\end{table}

\begin{figure}[h]
\begin{center}
 \includegraphics[scale=0.5]{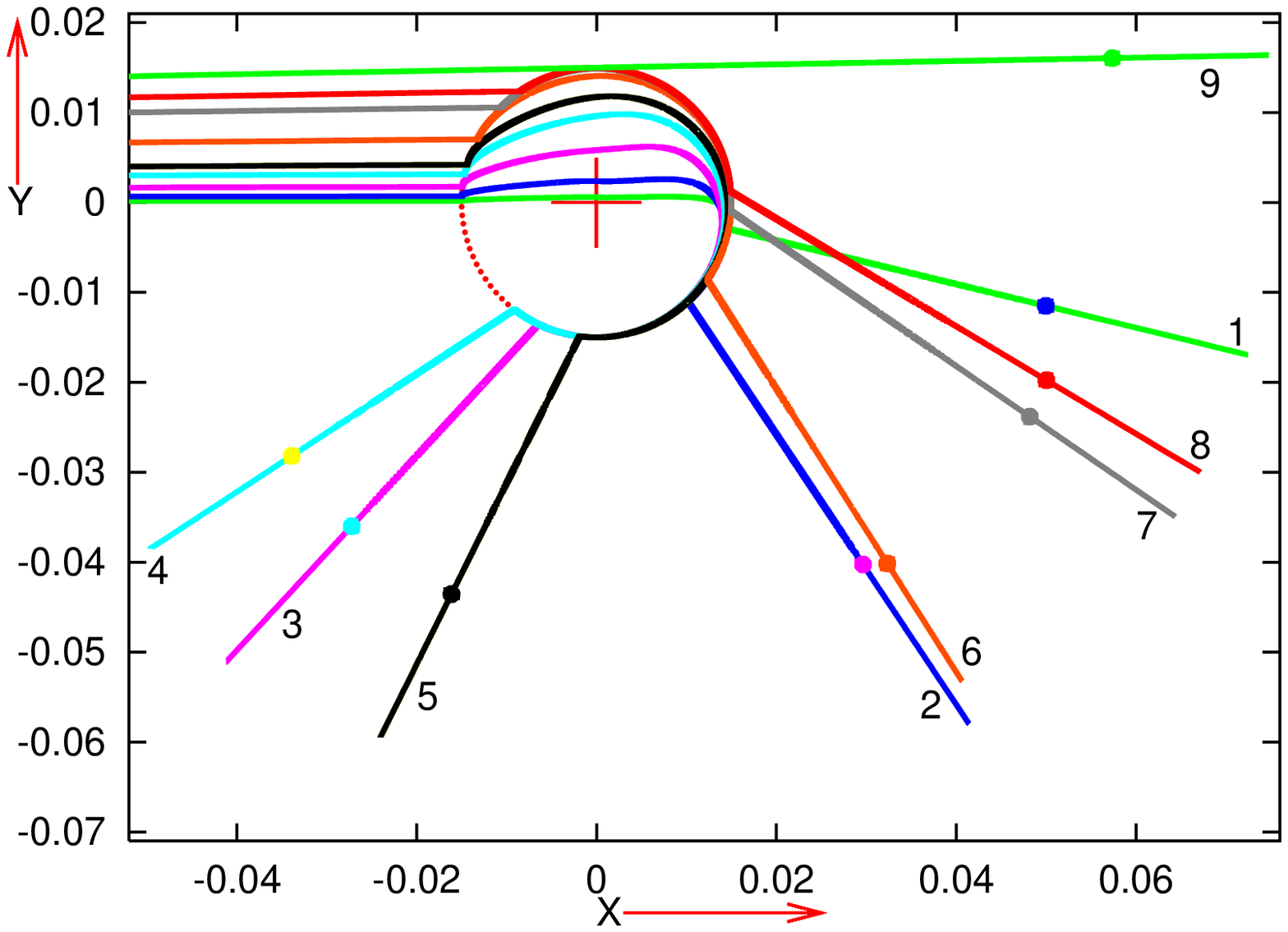}
${  }$ \\ [-2.0cm]
 \includegraphics[scale=0.5]{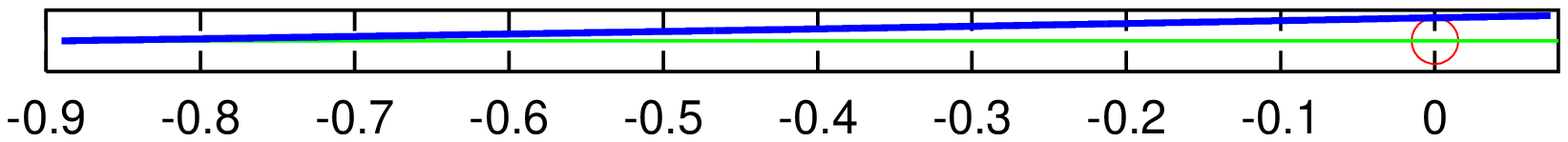}
\caption{{\bf Upper panel:} Projections of the rays listed in Table
\ref{obsIrays} on a surface of constant $t$ along the flow lines of the cosmic
dust. Observer I is at $X \approx -0.8898, Y = 0$ beyond the left margin of the
figure. The large dots mark the intersections of the rays with the
last-scattering hypersurface. The dotted circle has the radius $r = A_1 + B_1$,
where the BB hump has its edge. More explanation in the text. {\bf Lower panel:}
Ray 9 shown all the way between Observer I and the BB. }
 \label{noaxraysI}
\end{center}
\end{figure}

\begin{figure}[h]
\begin{center}
 \includegraphics[scale=0.5]{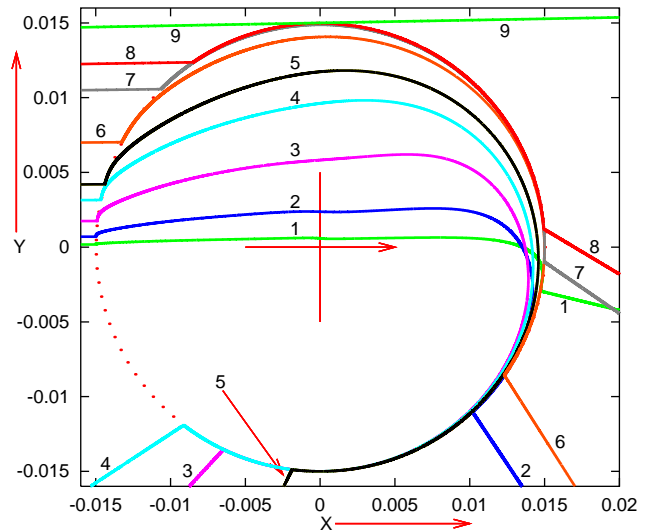}
\caption{The region near the BB hump in Fig. \ref{noaxraysI}. The arrow on the
horizontal arm of the cross points in the direction of the dipole maximum. More
explanation in the text.}
 \label{oIraysmall}
\end{center}
\end{figure}

In Figs. \ref{noaxraysI} and \ref{oIraysmall} the coordinates are
\begin{equation}\label{9.18}
X = - r \cos \vartheta, \qquad Y = r \sin \vartheta.
\end{equation}
Figure \ref{noaxraysI} shows the projections of the rays from Table
\ref{obsIrays} on a surface of constant $t$ along the flow lines of the dust in
a neighbourhood of the QSS region. Figure \ref{oIraysmall} is a closeup view on
the vicinity of the BB hump. The dotted circle is at $r = r_b$, the
$r$-coordinate of the edge of the BB hump. The cross marks the center $r = 0$ of
the dotted circle; the arrow on the horizontal arm of the cross in Fig.
\ref{oIraysmall} points in the direction of the Szekeres dipole maximum. The
large dots in Fig. \ref{noaxraysI} mark the points where the rays intersect the
LSH. The endpoints of the rays are where the numerical calculation determined
their crossing the BB. Figures \ref{noaxraysI} and \ref{oIraysmall} are nearly
the same as the corresponding ones for the L--T/Friedmann model in Ref.
\cite{Kras2016a}; there are only small quantitative differences between them.
They are shown here to facilitate comparisons with the images of the rays
reaching Observers II and III further on.

Ray 0 is not included in the figures because, at their scale, it would coincide
with the $Y = 0$ axis. It is included in the table in order to show how $1 +
z_{\rm LSH}$ abruptly jumps from the near-zero value (\ref{8.12}) on an axial
ray to a large positive value on a ray that is only slightly nonaxial.

The redshifts initially increase with the viewing angle. The maximum $z_{\rm
LSH}$ is achieved on Ray 8 inside the image of the source, not at its edge, and
it is larger than the background (\ref{9.17}). The same thing happened in the
L--T/Friedmann model \cite{Kras2016a}, and will again occur for Observers II and
III further in this paper. Ray 9 just grazes the world-tube $r = r_b$, and
$z_{\rm LSH}$ on it is close to (\ref{9.17}). Its $k^{\vartheta}_o$ was
determined by trial and error: for each ray the program that calculated its path
determined the minimum $r \df r_{cl}$ along it, and Ray 9 is the one where
$r_{cl} - r_b = 0.0000000000735095811$ was reasonably small.

The rays abruptly change their direction every time they come near to the
surface $r = r_b$. The change is sharper on the second intersection with $r =
r_b$ where the ray is closer to the BB. When the rays travel over the BB hump
further from its edge the deflections are smaller.

The angle of deflection depends on the interval of $t$ that the ray spends near
the edge of the BB hump. Ray 1 meets $r = r_b$ nearly head-on and does not
strongly change direction on first encounter. On second encounter, it is closer
to the BB and is forced to bend around more.

The other rays meet the $r = r_b$ surface at smaller angles than Ray 1, so they
stay near it for longer times. For Rays 3, 4 and 5, this causes a much stronger
deflection than for Ray 1. For Rays 6 -- 8, another effect prevails: they fly
farther from the axis, so they approach the BB at larger $t - t_B$ and stay over
it for a shorter time; therefore the deflection angle decreases again. Ray 9
does not enter the Szekeres region but only touches it, so it propagates almost
undisturbed as in the Friedmann region.

Figures \ref{noaxraysI} and \ref{oIraysmall} show only those rays for which
$k^{\vartheta}_o > 0$. The images of the rays with $k^{\vartheta}_o <0$ are
mirror reflections of those shown. In fact, since $\vartheta = 0$ is the axis of
symmetry, the image will be the same for every $\varphi$, so one should imagine
the complete collection of constant-$\varphi$ null geodesics by rotating Figs.
\ref{noaxraysI} and \ref{oIraysmall} around the $\vartheta = 0$ axis.

\subsection{Rays reaching Observer II}

\begin{figure}[h]
\begin{center}
 \includegraphics[scale=0.5]{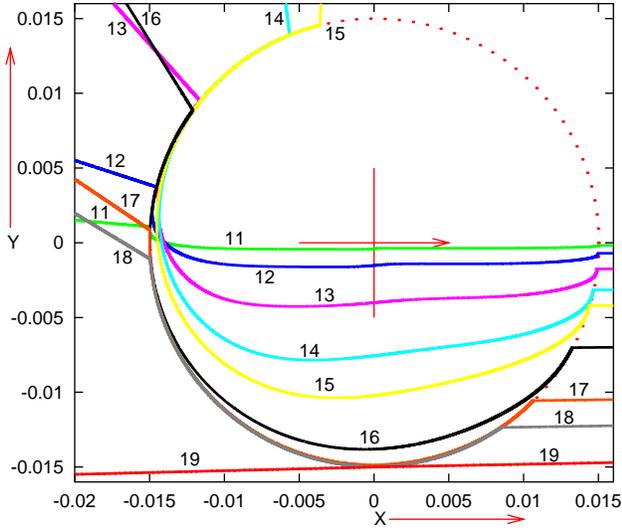}
\caption{The analogue of Fig. \ref{oIraysmall} for Observer II, who is at $X
\approx 0.889, Y = 0$ beyond the right margin of the figure. See Table
\ref{obsIIrays} for the parameters of the rays.}
 \label{oIIraysmall}
\end{center}
\end{figure}

Table \ref{obsIIrays} and Fig. \ref{oIIraysmall} are analogues of Table
\ref{obsIrays} and Fig. \ref{oIraysmall} for Observer II. The analogue of Ray
$n$ from Table \ref{obsIrays} is Ray $10 + n$ in Table \ref{obsIIrays}. The
$k^{\vartheta}_o$ are the same as in Table \ref{obsIrays}, with the exception of
Ray 19 -- see below for an explanation. The angular radii are slightly smaller
here because $\Phi_o$ for Observer II is slightly smaller than (\ref{9.11}):
\begin{equation}\label{9.19}
(\Phi_o)_{\rm obs\ 2} = 0.40181424093371831.
\end{equation}
But at the level of precision used in the tables, the angular radii for Rays 11
-- 18 are the same as those for Rays 1 -- 8. The analogue of Ray 0 is not
included.

\begin{table}[h]
\begin{center}
\caption{Parameters of rays reaching Observer II}
\bigskip
\begin{tabular}{|c|c|c|c|c|c|c|}
  \hline \hline
  Ray & $k^{\vartheta}_o$ & $1 + z$ at LSH \\
  \hline \hline
  11 & 0.0005 & 358.22989174485627 \\
  \hline
  12 & 0.002 & 388.80853980783820 \\
  \hline
  13 & 0.005 & 408.06476517495747 \\
  \hline
  14 & 0.009 & 504.79183448874682 \\
  \hline
  15 & 0.012 & 620.08511872418046 \\
  \hline
  16 & 0.02 & 885.02972357472424 \\
  \hline
  17 & 0.03 & 970.70644144723383 \\
  \hline
  18 & 0.035 & 982.13817446295479 \\
  \hline
  19 & 0.04205 & 951.83804564661989 \\
  \hline \hline
\end{tabular} \\
 \label{obsIIrays}
\end{center}
\end{table}

Ray 19 grazes the edge of the Szekeres region -- so its $k^{\vartheta}_0$
determines the angular radius of the whole source by (\ref{9.16}). Since $r_{\rm
obs}$ is smaller here, the angular radius for Ray 19 is larger than for Ray 9;
it is
\begin{equation}\label{9.20}
\alpha_{\rm II} = 0.9681^{\circ}.
\end{equation}

The values of $1 + z_{\rm LSH}$ in Table \ref{obsIIrays} are different from
those in Table \ref{obsIrays}, but the general pattern is the same: $z_{\rm
LSH}$ initially increases with the viewing angle, achieves a maximum inside the
image of the source, then decreases to the background value at its edge. The
maximum is achieved at the same $k^{\vartheta}_o$ as before, on Ray 18.

\subsection{Rays reaching Observer III}

\begin{figure}[h]
\begin{center}
 \includegraphics[scale=0.5]{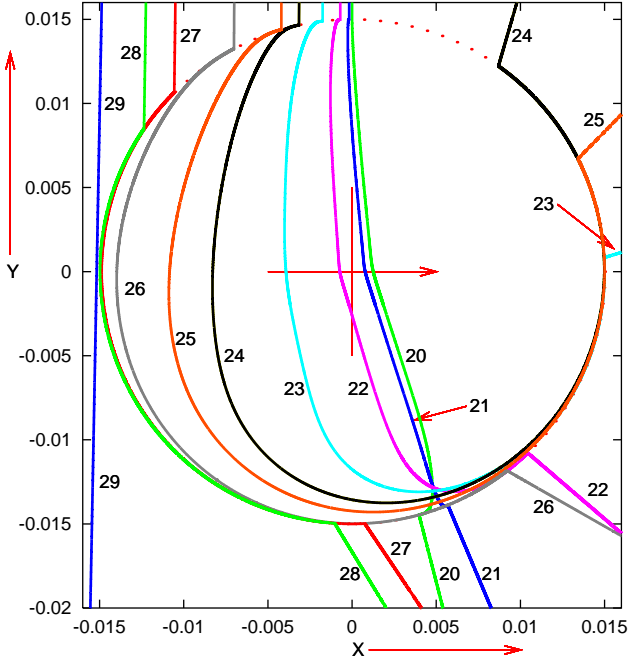}
\caption{The analogue of Fig. \ref{oIraysmall} for Observer III, who is at $X =
0, Y \approx 0.889$ above the upper edge of the figure. Only the rays with
$k^{\vartheta}_o \leq 0$ are shown; see Table \ref{obsIIIraysleft} for their
parameters.}
 \label{oIIIraysmleft}
\end{center}
\end{figure}

\begin{figure}[h]
\begin{center}
 \includegraphics[scale=0.5]{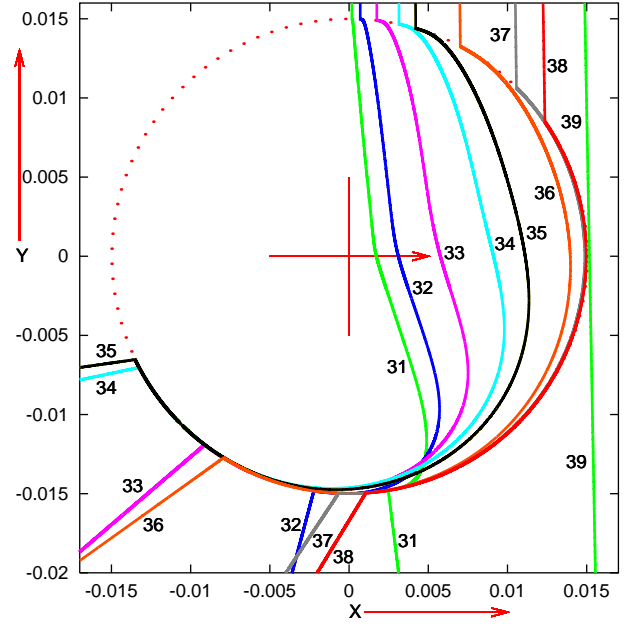}
\caption{The rays that have $k^{\vartheta}_o > 0$ at Observer III; see Table
\ref{obsIIIraysright} for their parameters.}
 \label{oIIIraysmright}
\end{center}
\end{figure}

Observer III, unlike Observers I and II, is not located on the axis of symmetry,
so the (past-directed) rays going off from her position with $k^{\vartheta}_o <
0$ will not be mirror images of those with $k^{\vartheta}_o > 0$. Therefore,
these two groups of rays are shown in separate tables and separate figures.
Table \ref{obsIIIraysleft} and Fig. \ref{oIIIraysmleft} contain the rays for
which $k^{\vartheta}_o \leq 0$; the rays in Table \ref{obsIIIraysright} and Fig.
\ref{oIIIraysmright} have $k^{\vartheta}_o > 0$. The set of values of
$\left|k^{\vartheta}_o\right|$ is the same as in Table \ref{obsIrays} and Fig.
\ref{oIraysmall}. The analogues of Ray $n$ from Table \ref{obsIrays} are Ray $20
+ n$ in Table \ref{obsIIIraysleft} and Ray $30 + n$ in Table
\ref{obsIIIraysright}.

\begin{table}[h]
\begin{center}
\caption{Parameters of rays with $k^{\vartheta}_o \leq 0$ at Observer III}
\bigskip
\begin{tabular}{|c|c|c|c|c|c|c|}
  \hline \hline
  Ray & $k^{\vartheta}_o$ & $1 + z$ at LSH \\
  \hline \hline
  20 & 0.0 & 342.29964855437106 \\
  \hline
  21 & -0.0005 & 350.64558337051187 \\
  \hline
  22 & -0.002 & 337.49308380652388 \\
  \hline
  23 & -0.005 & 361.19113331483726 \\
  \hline
  24 & -0.009 & 470.62189702521152 \\
  \hline
  25 & -0.012 & 629.72937110236398 \\
  \hline
  26 & -0.02 & 900.56138279350250 \\
  \hline
  27 & -0.03 & 971.41838000807513 \\
  \hline
  28 & -0.035 & 982.30363263812137 \\
  \hline
  29 & -0.0425 & 951.83650139022654 \\
  \hline \hline
\end{tabular} \\
 \label{obsIIIraysleft}
\end{center}
\end{table}

The value of $\Phi_o$ here is between the previous ones,
\begin{equation}\label{9.21}
(\Phi_o)_{\rm obs\ 3} = 0.40192128311507536,
\end{equation}
while $t_o = 0$ does not differ significantly from (\ref{8.20}) and
(\ref{8.21}), so the angular radii would also be intermediate; they are not
listed in the tables.

The most conspicuous difference from the previous cases is in Ray 20, which
proceeds along $\vartheta = \pi/2$ in the Friedmann region: it is deflected
toward larger $\vartheta$ on entry to the Szekeres region, and bends oppositely
to all other rays on leaving it. Rays 21 and 22 get deflected so strongly that
they cross the line $\vartheta = \pi/2, 3 \pi/2$ well inside the Szekeres
region, unlike their analogues, Rays 1, 2, 11 and 12, which cross the $\vartheta
= 0, \pi$ lines just before leaving the Szekeres region. Beginning with Ray 23,
the paths of the rays become similar (though different in numerical detail) to
the corresponding ones for Observers I and II.

The pattern of $1 + z_{\rm LSH}$ across the image of the source here is
different from those for Observers I and II: with decreasing $k^{\vartheta}_o <
0$ the redshift achieves a minimum on Ray 22, then a maximum larger than in the
background on Ray 28; it then drops to the background value. One ray in this
family (not shown) will pass through $r = 0$, but with $0 \neq \vartheta \neq
\pi$, so it will not have $z = -1$ at the BB for the reason indicated under Eq.
(\ref{6.4}). See also Ref. \cite{Kras2016b}, where rays passing through $r = 0$
were numerically integrated for the same kind of Szekeres dipole (but with a
different BB profile and with $a^2 = 0.1$) -- only those proceeding along
$\vartheta = 0, \pi$ had $z \approx -1$ near the BB.

\begin{table}[h]
\begin{center}
\caption{Parameters of rays with $k^{\vartheta}_o > 0$ at Observer III}
\bigskip
\begin{tabular}{|c|c|c|c|c|c|c|}
  \hline \hline
  Ray & $k^{\vartheta}_o$ & $1 + z$ at LSH \\
  \hline \hline
  31 & 0.0005 & 360.79504513233314 \\
  \hline
  32 & 0.002 & 381.93470986017479 \\
  \hline
  33 & 0.005 & 453.34271919911635 \\
  \hline
  34 & 0.009 & 576.90432434248658 \\
  \hline
  35 & 0.012 & 682.52857109479601 \\
  \hline
  36 & 0.02 & 895.45677016306377 \\
  \hline
  37 & 0.03 & 970.72628947084468 \\
  \hline
  38 & 0.035 & 982.16761884746518 \\
  \hline
  39 & 0.0425 & 951.83650139022654 \\
  \hline \hline
\end{tabular} \\
 \label{obsIIIraysright}
\end{center}
\end{table}

For rays with $k^{\vartheta}_o > 0$ the pattern of $1 + z_{\rm LSH}$ is similar
to that for Observer II: there is only the maximum, on Ray 38. However, the
values of $1 + z_{\rm LSH}$ differ, some of them substantially, from their
counterparts in Table \ref{obsIIrays}.

The paths of the rays are similar to those for Observers I and II, but the angle
of deflection is smaller for each ray here. Also, the rays bend away from the $X
= 0$ axis near the $Y = 0$ line -- this effect was not visible for Observer I
and barely noticeable for Observer II.

\section{Redshift profiles along nonaxial null geodesics}\label{noaxreds}

\setcounter{equation}{0}

\begin{figure}[h]
\begin{center}
 \includegraphics[scale=0.5]{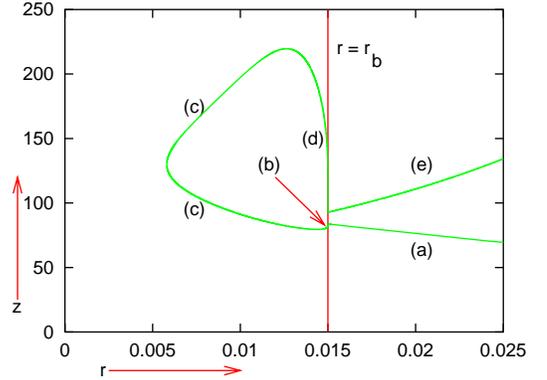}
\caption{The $z(r)$ relation for Ray 3 from Fig. \ref{noaxraysI} and its
characteristic branches. Observer I is at $r \approx 0.8898$ beyond the right
edge of this figure. See text for explanation.}
 \label{redprofI}
\end{center}
\end{figure}

\begin{figure}[h]
\begin{center}
 \includegraphics[scale=0.6]{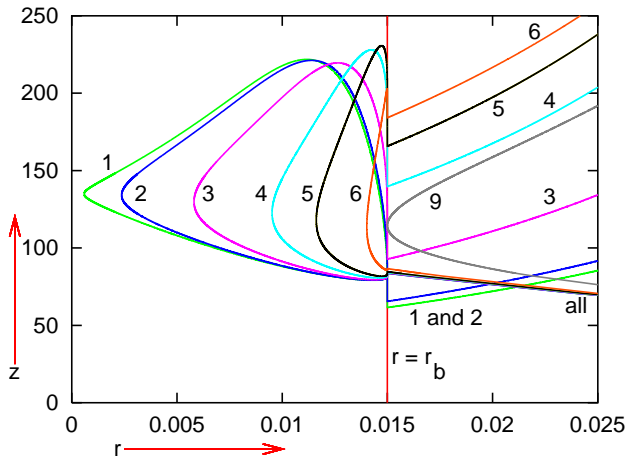}
\caption{The $z(r)$ relations for Rays 1 -- 6 and 9 from Fig. \ref{noaxraysI}.
Along Rays 1 -- 6, $z$ has two maxima and two minima, which points to the
existence of an analogue of the Extremum Redshift Surface along them. Ray 9 does
not enter the blueshift-generating region, so $z$ is increasing all along it.}
\label{redprofsI}
\end{center}
\end{figure}

The $z$-profiles along Rays 1 -- 6 and 9 are shown in Figs. \ref{redprofI} and
\ref{redprofsI}; they are similar to those in the L--T/Friedmann model
\cite{Kras2016a}. They show that analogues of the ERS (call them ERS') exist
also along nonaxial rays. Figure \ref{redprofI} shows the $z(r)$ relation for
Ray 3 in a neighbourhood of $r = r_b$; it is a key to reading Fig.
\ref{redprofsI}. In segment (a) of the ray, $z$ increases from 0 at the observer
to a local maximum at $r \approx r_b$, where the (past-directed) ray intersects
the outer branch of the ERS' for the first time. Then, in segment (b), $z$
decreases to a local minimum at a slightly smaller $r$, where the ray intersects
the inner branch of the ERS' for the first time. Further along the ray, in
segment (c), $z$ increases until it reaches the second local maximum at the
second intersection of the ray with the inner branch of the ERS'. Then, in
segment (d), $z$ decreases up to the second intersection of the ray with the
outer branch of the ERS', where it achieves its second and last local minimum.
From then on, in segment (e), $z$ keeps increasing up to $\infty$ achieved at
the BB.

Along Rays 1 and 2 in Fig. \ref{redprofsI}, the second minimum of $z$ is smaller
than the first maximum, so those $z(r)$ curves self-intersect.

\section{Fitting the radiation sources in the celestial sphere}\label{inthesky}

\setcounter{equation}{0}

Imagine a radiation source to be a disk on the celestial sphere of angular
radius $\vartheta_0$. How many such disks would fit into the celestial sphere at
the same time?

An equivalent question is, how many non-overlapping circles of a given radius
can be drawn on a sphere of a given radius? A rough answer would be obtained by
dividing the surface area of the sphere by the surface area inside the circle.
But this would be an overestimate -- the circles cannot cover the sphere
completely. A better approximation is to inscribe each circle into a quadrangle
of arcs of great circles on the sphere. Such figures cannot cover the sphere
either, but this method takes into account some of the area outside the circles.
Details of the calculation are presented in the Appendix. The area of the sphere
divided by the area of the quadrangle is
\begin{equation}\label{11.1}
{\cal N} = \frac {\pi} {\arcsin \left(\sin^2 \vartheta_0\right)}.
\end{equation}
Taking $\vartheta_0 = 0.5^{\circ}$, the current resolution of the GRB detectors
(see footnote \ref{Linda}), we obtain
\begin{equation}\label{11.2}
{\cal N}_{0.5} \approx 41,254.
\end{equation}
With $\vartheta_0 = 0.96767^{\circ}$ of Table \ref{obsIrays}, we obtain
\begin{equation}\label{11.3}
{\cal N}_{0.96767} \approx 11,014.
\end{equation}
Finally, with $\vartheta_0 = 0.9681^{\circ}$, as in (\ref{9.20}), we obtain
\begin{equation}\label{11.4}
{\cal N}_{0.9681} \approx 11,005.
\end{equation}

It is instructive to compare these numbers with the number of GRBs detected in
observations. This author was not able to get access to a definitive answer, but
here is an estimate based on partial information. The BATSE (Burst and Transient
Source Explorer) detector, which worked in the years 1991 -- 2000, discovered
2704 GRBs \cite{BATSE} (it was de-orbited in 2000 \cite{deorbit}). Assuming the
same rate of new discoveries, 8112 GRBs should have been detected between 1991
and now -- still fewer than (\ref{11.4}).

When the angular radius is divided by $f$, the number of possible sources in the
sky should be multiplied by $f^2$. Equation (\ref{11.1}) approximately confirms
this, since for small $\vartheta_0$ we have $\sin \vartheta_0 \approx
\vartheta_0 \approx \arcsin \vartheta_0$.)

\section{Possible and necessary improvements of the model}\label{impro}

\setcounter{equation}{0}

The model presented here accounts for the lowest frequency of the radiation in
the observed GRBs (the model of highest-frequency GRBs was discussed in Ref.
\cite{Kras2016a}). The angular radius of the radiation sources seen by the
present observer is twice as large as the current observations allow (nearly
$1^{\circ}$ in the model vs. $0.5^{\circ}$ -- the resolution of the GRB
detectors; see footnote \ref{Linda}). In order to decrease this angle, the BB
hump that emits the radiation should be made narrower or lower; in the second
case it would be further away from the observer seeing the high-frequency flash.
The BB profile chosen in this paper cannot be the limit of improvement. The
first attempt to explain the GRBs using a cosmological blueshift resulted in a
model \cite{Krasfail} whose hump had the height $A_0 + B_0 = 0.026$ NTU and
width $A_1 + B_1 = 0.108$. By experimenting with the parameters of the hump, the
numbers in (\ref{3.8}) were achieved; i.e. the height was decreased $\approx
206$ times and the width 7.2 times. The result of such a blind search cannot be
the best possible. In particular, other classes of shapes of the BB hump should
be tried.

To get small $1 + z$, the BB profile should be such that the blueshifted ray
spends as much time as possible traveling above the LSH but below the ERS. As
follows from (\ref{7.7}) and (\ref{7.8}), the room under the ERS becomes larger
when $\dril {t_B} r$ is larger and when $a$ is smaller. The problem with small
$a$ was described in Sec. \ref{corrM2}, but it might be overcome using a greater
numerical precision. A larger $\dril {t_B} r$ tends to make the BB hump higher.
In order to keep the hump acceptably low, the large $\dril {t_B} r$ has to be
limited to a short interval of $r$ -- this is where the steep slope of the hump
in Fig. \ref{smallz} came from.

A serious limitation is the fact mentioned in Sec. \ref{ERS} that the ERS is
tangent to the BB at $r = 0$. If this could be overcome, the rays would stay in
the blueshift-generating region (below the ERS) for a longer time interval, and
so the required $1 + z$ range could be achieved with a lower or narrower hump.

Further optimizations are possible. For example, the function $E(r)$ here has
the Friedmann shape (\ref{3.2}) throughout the Szekeres region -- obviously one
should check what happens when it has other shapes. Friedmann backgrounds other
than the one of Sec. \ref{backgrmodel} should be tested. Szekeres dipoles other
than (\ref{3.9}) should also be tested, in particular non-axially-symmetric
ones. Carrying out such tests is laborious -- it involves finding, by numerical
shooting, the minimum of a function of several variables (in this paper these
were 7 variables: the five in (\ref{3.8}), the $a$ of (\ref{3.9}) and the
$\Delta t_c$ of (\ref{8.4})).

Similar to the L--T model of Ref. \cite{Kras2016a}, the model presented here
implies too-long durations for the high-frequency flashes and for their
afterglows. This is because, in axially symmetric models, once the observer and
the source are placed on the symmetry axis, they stay there forever -- the
source does not drift \cite{KrBo2011,QABC2012,KoKo2017}. The only changes of the
observed frequency and intensity may then occur because the observer receives
rays emitted from different points of the BB hump along the same line of sight,
so the changes occur on the cosmological time scale and are much slower than in
the observed GRBs (see Ref. \cite{Kras2016a} for the numbers).

A nonsymmetric Szekeres model offers a new possibility. In such a model there
also exist two opposite directions along which radiation is strongly blueshifted
\cite{Kras2016b}. However, the cosmic drift \cite{KrBo2011,QABC2012,KoKo2017}
will cause an observer who was initially in the path of one of those preferred
rays to be off it after a while. The time scale of this process should be short,
as a consequence of the very large distance between the source and the observer
and of the discontinuous change from blueshift to redshift as soon as the
strongly blueshifted ray misses the observer.

One solution of the duration problem has already been tested, and will be
submitted for publication soon. If there is another QSS region between the
radiation source and the observer, then the cosmic drift in the intervening QSS
region will cause the highest-frequency ray to miss the observer after 10
minutes or less. This satisfactorily solves the problem of the duration of the
high-frequency flash, but not the problem of the duration of the afterglow. The
latter still awaits solution.

\section{Summary and conclusions}\label{sumup}

\setcounter{equation}{0}

In Ref. \cite{Kras2016b}, existence and properties of blueshifts in exemplary
simple quasispherical Szekeres models were investigated. Using that knowledge,
in the present paper it was investigated whether a QSS mass dipole superposed on
a L--T background would allow better mimicking of gamma-ray bursts by
cosmological blueshifting than in Ref. \cite{Kras2016a}, where pure L--T models
were used.

The axially symmetric QSS model was introduced in Secs. \ref{QSSS} and
\ref{QSShere}. The QSS region is matched to a negative-spatial-curvature
Friedmann background (Sec. \ref{backgrmodel}), chosen for correspondence with
earlier papers by this author \cite{Kras2014d,Kras2014a}. After presenting
definitions and preliminary information in Secs. \ref{symmetric}, \ref{axreds}
and \ref{ERS}, in Sec. \ref{corrM2} the parameters of the QSS model are chosen
such that at present the highest frequency of the blueshifted radiation agrees
with the lowest frequency of the observed GRBs (this agreement requires that the
blueshift between the last scattering and the present time obeys $1 + z \leq
1.689 \times 10^{-5}$ \cite{Kras2016a}). The introduction of the Szekeres dipole
has the consequence that the required $1 + z$ is achieved with a lower hump in
the BB profile, which is thus at a greater distance from the observer than in
the L--T model. In Sec. \ref{noax}, the paths of nonaxial light rays reaching
three different present observers are presented. The observers are placed in
prolongation of the mass-dipole maximum axis, of the dipole minimum axis, and of
the dipole equator. The distributions of the observed redshift across the image
of the source are different for each observer, and the angular radii of the
source are between $0.96767^{\circ}$ and $0.9681^{\circ}$. This is nearly twice
as much as the current GRB observations allow, but the model has the potential
to be improved (see Sec. \ref{impro}). In Sec. \ref{noaxreds}, the redshift
profiles {\em along} nonaxial rays were calculated in order to show that extrema
of redshift also exist along them. In Sec. \ref{inthesky} it was estimated that
with the angular radii of the radiation sources being between $0.96767^{\circ}$
and $0.9681^{\circ}$, approximately 11,000 such sources could be simultaneously
fitted into the sky of the present observer. Finally, possible further
improvements in the model were discussed in Sec. \ref{impro}.

The models of generating the high-frequency radiation flashes discussed here and
in Ref. \cite{Kras2016a} are subject to two kinds of tests:

1. In the future, the observers should be able to resolve the fuzzy disks they
now see as GRB sources (see footnote \ref{Linda}), and measure the distribution
of radiation frequencies and intensities across them. Then it will be possible
to compare those distributions with model predictions. A model that would
predict such a distribution correctly could then be used to get information
about the sources.

2. If the gamma flashes are generated simultaneously with the CMB radiation, as
proposed here and in Ref. \cite{Kras2016a}, then they are observed now as
short-lived because its source comes into and out of the observer's view, but
has existed there since the last-scattering epoch. In this case, the central
high-frequency ray should be surrounded by rays with positive redshifts smoothly
blending with the CMB background at the edge of the source image, as shown in
tables in Sec. \ref{noax}. But if a source of the radiation flash lies later
than the last scattering, then it is independent of the CMB. It should black out
all CMB rays within some angle around the central ray, and the redshift profile
across the image of the source would not need to continuously match the CMB at
the edge.

This author does not wish to question the validity of the GRB models proposed so
far. The motivation for this work was this: history of science teaches us that
if a well-tested theory predicts a phenomenon, then the prediction has to be
taken seriously and checked against experiments and observations. Since general
relativity clearly predicts that some of the light generated during last
scattering might reach us with strong blueshift, consequences of this prediction
have to be worked out and submitted to tests. In trying to accommodate
blueshifts, the suspicion fell on the GRBs because it is generally agreed that
at least some of their sources lie billions of years to the past from now
\cite{Perlwww}. The BB humps discussed here would lie about twice as far, at
$\approx 13.6$ Gyr to the past, by (\ref{3.6}). For the relativity theory, it
would be interesting to know whether at least some of the observed GRBs are
powered by the mechanism discussed here.

\appendix

\section{How many circles of a given radius can be drawn on a sphere of a
given radius?} \label{circles}

\setcounter{equation}{0}

\begin{figure}[h]
\begin{center}
\includegraphics[scale=0.5]{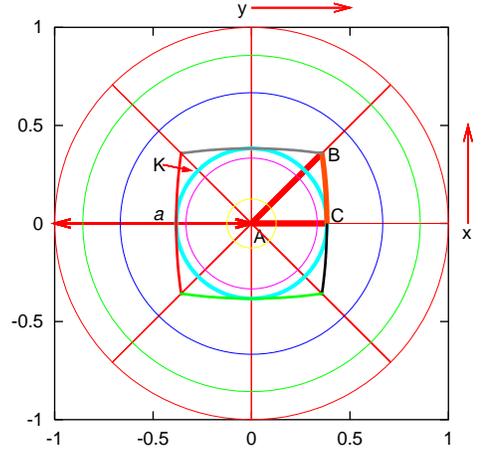}
\caption{View from the $z > 0$ axis on the cone and the pyramid. They intersect
the sphere along the circle K and the curvilinear quadrangle shown in thicker
lines, respectively.}
 \label{curvsquareup}
\end{center}
\end{figure}

\begin{figure}[h]
\begin{center}
\includegraphics[scale=0.5]{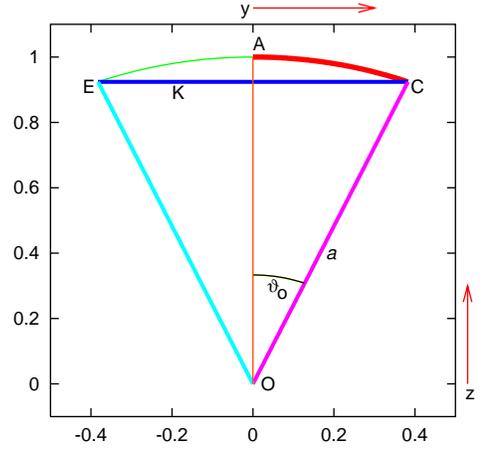}
\caption{Projection of the cone and of the pyramid from Fig. \ref{curvsquareup}
on the $x = 0$ coordinate plane. The lines OE and OC are intersections of the
cone (and of the faces of the pyramid) with the plane of the figure; the letters
K, A and C have the same meaning as in Fig. \ref{curvsquareup}.}
\label{curvsquareside}
\end{center}
\end{figure}

\begin{figure}
\begin{center}
\includegraphics[scale=0.5]{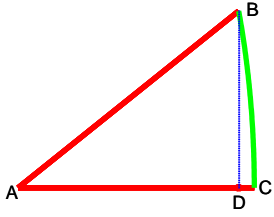}
\caption{A sketch to calculate the surface area of the triangle ABC on a sphere.
See text.}
 \label{curvtriangle}
\end{center}
\end{figure}

Imagine a circle K drawn on a sphere S of radius $a$ and a cone that intersects
S along K and has its vertex at the center of S; see Figs. \ref{curvsquareup}
and \ref{curvsquareside}. Let the opening angle of the cone be $\vartheta_0$.
Now imagine a square pyramid circumscribed on this cone. The pyramid intersects
S along the curvilinear quadrangle shown in thicker lines in Fig.
\ref{curvsquareup}. The part of S inside the quadrangle has the surface area 8
times the surface area inside the curvilinear triangle ABC; see also Fig.
\ref{curvtriangle}.

Suppose the center of the sphere is at $x = y = z = 0$, so the equation of the
sphere is $x^2 + y^2 + z^2 = a^2$, and the axis of the cone goes along the $z$
axis. The metric of the sphere in the $(x, y)$ coordinates is
\begin{eqnarray}\label{a.1}
&& {\rm d} x^2 + {\rm d} y^2 + {\rm d} z^2 = \\
&& \frac {\left(a^2 - y^2\right) {\rm d} x^2 + 2xy {\rm d} x {\rm d} y +
\left(a^2 - x^2\right) {\rm d} y^2} {a^2 - x^2 - y^2}, \nonumber
\end{eqnarray}
and so the surface element of the sphere is
\begin{equation}\label{a.2}
\sqrt {g} {\rm d} x {\rm d} y = \frac a {\sqrt{a^2 - x^2 - y^2}} {\rm d} x {\rm
d} y.
\end{equation}
The side AC of the triangle lies in the plane $x = 0$, and $y$ on it changes
from 0 to $a \sin \vartheta_0$. The side AB lies in the plane $y = x$. The
$y$-coordinate of the point B is
\begin{equation}\label{a.3}
y_{\rm B} = \frac {a \sin \vartheta_0} {\sqrt{1 + \sin^2\vartheta_0}},
\end{equation}
as is easy to calculate knowing that this point lies simultaneously on the
sphere $x^2 + y^2 + z^2 = a^2$, in the plane $y = x$ and in the plane $z = y
\cot \vartheta_0$ that contains the right face of the pyramid. The auxiliary
point D has the same $y$-coordinate as B. The arc BC (which is part of the
intersection of the right face of the pyramid with the sphere) obeys the
equation
\begin{equation}\label{a.4}
x = \sqrt{a^2 - \frac {y^2} {\sin^2\vartheta_0}} \df x_{\rm BC}(y).
\end{equation}
The surface area of the triangle ABC is thus
\begin{eqnarray}
&& S_{\rm ABC} = \int_0^{y_{\rm B}} {\rm d} y \int_0^y \frac a {\sqrt{a^2 - x^2
- y^2}}\ {\rm d} x \nonumber \\
&& + \int_{y_{\rm B}}^{a \sin \vartheta_0} {\rm d} y \int_0^{x_{\rm BC}(y)}
\frac a {\sqrt{a^2 - x^2 - y^2}}\ {\rm d} x \label{a.5} \\
&& = \int_0^{y_{\rm B}} a \arcsin \left(\frac y {\sqrt{a^2 - y^2}}\right) {\rm
d} y \nonumber \\
&& + \int_{y_{\rm B}}^{a \sin \vartheta_0} a \arcsin \left(\frac {x_{\rm BC}(y)}
{\sqrt{a^2 - y^2}}\right) {\rm d} y. \label{a.6}
\end{eqnarray}

The two integrals in (\ref{a.6}) are
\begin{eqnarray}
S_I &=& \frac {a^2 \vartheta_0 \sin \vartheta_0} {\sqrt{1 + \sin^2\vartheta_0}}
- \tfrac 1 2\ a^2 \arcsin \left(\sin^2 \vartheta_0\right), \label{a.7} \\
S_{II} &=& - \frac {a^2 \vartheta_0 \sin \vartheta_0} {\sqrt{1 +
\sin^2\vartheta_0}} + a^2 \arcsin \left(\sin^2 \vartheta_0\right). \label{a.8}
\end{eqnarray}
So, the area of the triangle ABC is $\tfrac 1 2\ a^2 \arcsin \left(\sin^2
\vartheta_0\right)$, and the area of the quadrangle in Fig. \ref{curvsquareup}
is
\begin{equation}\label{a.9}
S_{\rm quad} = 4 a^2 \arcsin \left(\sin^2 \vartheta_0\right).
\end{equation}
(When $\vartheta_0 = \pi/2$, this gives the obvious result $2 \pi a^2$.)

{\small

Hints for the less-trivial parts of calculating the integrals:

In $S_I$ change the variables by $\arcsin \left(\frac y {\sqrt{a^2 -
y^2}}\right) = w$ and integrate by parts to get rid of the factor $w$ under the
integral.

In $S_{II}$ change the variables by $y = a \sin \vartheta_0 \sin u$, then
integrate by parts to get rid of $\arcsin$ under the integral, and finally use
the identity $\arctan \lambda = \arcsin \left(\frac {\lambda} {\sqrt{1 +
\lambda^2}}\right)$.

         }

Now an approximate answer to the question in the title can be given. The
quadrangles will not cover the whole surface of the sphere, but by dividing the
surface area of the sphere, $4 \pi a^2$, by $S_{\rm quad}$, we obtain an upper
bound on the number of nonoverlapping circles that can be drawn on the sphere;
it is (\ref{11.1}).

\begin{acknowledgments}
In deriving the geodesic equations, the computer-algebra system Ortocartan
\cite{Kras2001,KrPe2000} was used.
\end{acknowledgments}

\end{document}